\documentclass[twocolumn]{aastex62}
\usepackage{amsmath}
\usepackage{url}
\usepackage{lineno}

\graphicspath{{./}{figures/}}
\shorttitle{THOR}
\shortauthors{Moeyens et al.}

\begin{document}

\title{THOR: An Algorithm for Cadence-Independent Asteroid Discovery}

\correspondingauthor{Joachim Moeyens}
\email{moeyensj@uw.edu}

\author[0000-0001-5820-3925]{Joachim Moeyens}
\altaffiliation{B612 Asteroid Institute Fellow}
\affiliation{Department of Astronomy and the DIRAC Institute, University of Washington, 3910 15th Avenue NE, Seattle, WA 98195, USA}

\author[0000-0003-1996-9252]{Mario Juri\'{c}}
\affiliation{Department of Astronomy and the DIRAC Institute, University of Washington, 3910 15th Avenue NE, Seattle, WA 98195, USA}

\author{Jes Ford}
\affiliation{Department of Astronomy and the DIRAC Institute, University of Washington, 3910 15th Avenue NE, Seattle, WA 98195, USA}

\author[0000-0002-1312-5529]{Dino Bekte\v{s}evi\'{c}}
\affiliation{Department of Astronomy and the DIRAC Institute, University of Washington, 3910 15th Avenue NE, Seattle, WA 98195, USA}

\author[0000-0001-5576-8189]{Andrew J. Connolly}
\affiliation{Department of Astronomy and the DIRAC Institute, University of Washington, 3910 15th Avenue NE, Seattle, WA 98195, USA}

\author[0000-0002-1398-6302]{Siegfried Eggl}
\affiliation{Department of Astronomy and the DIRAC Institute, University of Washington, 3910 15th Avenue NE, Seattle, WA 98195, USA}

\author[0000-0001-5250-2633]{\v{Z}eljko Ivezi\'{c}}
\affiliation{Department of Astronomy and the DIRAC Institute, University of Washington, 3910 15th Avenue NE, Seattle, WA 98195, USA}

\author[0000-0001-5916-0031]{R. Lynne Jones} 
\affiliation{Department of Astronomy and the DIRAC Institute, University of Washington, 3910 15th Avenue NE, Seattle, WA 98195, USA}

\author[0000-0002-6825-5283]{J. Bryce Kalmbach}
\affiliation{Department of Astronomy and the DIRAC Institute, University of Washington, 3910 15th Avenue NE, Seattle, WA 98195, USA}

\author[0000-0002-7895-4344]{Hayden Smotherman}
\affiliation{Department of Astronomy and the DIRAC Institute, University of Washington, 3910 15th Avenue NE, Seattle, WA 98195, USA}

\begin{abstract}
We present ``Tracklet-less Heliocentric Orbit Recovery" (THOR), an algorithm for linking of observations of Solar System objects across multiple epochs that does not require intra-night tracklets or a predefined cadence of observations within a search window. By sparsely covering regions of interest in the phase space with ``test orbits", transforming nearby observations over a few nights into the co-rotating frame of the test orbit at each epoch, and then performing a generalized Hough transform on the transformed detections followed by orbit determination (OD) filtering, candidate clusters of observations belonging to the same objects can be recovered at moderate computational cost and little to no constraints on cadence. We validate the effectiveness of this approach by running on simulations as well as on real data from the Zwicky Transient Facility (ZTF). Applied to a short, 2-week, slice of ZTF observations, we demonstrate THOR can recover 97.4\% of all previously known and discoverable objects in the targeted ($a > 1.7$ au) population with 5 or more observations and with purity between 97.7\% and 100\%. This includes 10 likely new discoveries, and a recovery of an $e \sim 1$ comet C/2018 U1 (the comet would have been a ZTF discovery had THOR been running in 2018 when the data were taken). The THOR package and demo Jupyter notebooks are open source and available at \url{https://github.com/moeyensj/thor}.
\end{abstract}


\section{Introduction} \label{sec:intro}
The number of Solar System minor planet discoveries is growing rapidly thanks to the continuation of present day surveys such as Pan-STARRS \citep{Denneau2013} and the Catalina Sky Survey \citep{CSS2003}, and upcoming surveys such as the Vera C. Rubin Observatory's Legacy Survey of Space and Time (LSST) \citep{LSSTOverview2019} and NEOCam, recently renamed to NEO Surveyor \citep{MainzerNEOCam2016}. In about a
decade, the number of known objects will grow from the currently known 1 million to about 6 million minor planets. Such an increase in discoveries will enable a higher resolution look into the dynamical evolution of our Solar System. However, identifying minor planets in survey images and linking their detections into orbits continues to be a challenging problem. First, linking asteroid detections across multiple nights is difficult due to the sheer number of possible linkages, made even more challenging by the presence of false positives \citep{Kubica2007, Denneau2013, Veres2017, Veres2017a,  Jones2018}. Second, the motion of the observer makes the linking problem non-linear as minor planets will exhibit higher order motion on the topocentric sky over the course of weeks \citep{Holman2018}. Finally, once potential linkages have been established, they need to be confirmed as possible orbits using computationally expensive orbit determination software. For example, the Vera C. Rubin Observatory estimates it will discover nearly six million Main Belt asteroids that will be observed hundreds of times over the course of its ten year survey. Naively attempting to link hundreds of millions of asteroid detections over a ten year period is not computationally feasible \citep{Jones2018}. 

To make the linking problem more computationally tractable, surveys which aim to discover minor planets focus on constructing ``tracklets":
two dimensional sky-plane motion vectors consisting of two or more detections spaced typically 20-90 minutes apart that constrain the direction and rate of motion of potential moving objects. Tracklets are constructed to reduce the number of possible linkages that could be formed by providing information on  plausible direction and sky-plane angular velocity \citep{Kubica2007}. They are then linked into inter-night linkages known as ``tracks": sky-plane paths of motion containing several tracklets spanning up to $\sim15$ nights, typically modelled with low-order polynomials\footnote{More modern variants such as HelioLinC \citep{Holman2018} use transformations to the heliocentric coordinate system to linearize the linking problem.}. In the case of the LSST, for a moving object to be discoverable it must be observed at least twice a night on at least three unique nights within the $15$ day window to go through the tracklet-to-track creation process \citep{Jones2018}. In part due to the relative motion of the observer and the rate of motion of moving objects, both tracklets and tracks can exhibit high residuals relative to the fitted low-order polynomial, requiring relaxed fitting tolerances that can in turn lead to the creation of many spurious candidate linkages. Orbit determination (OD) algorithms are therefore required to run on each candidate track so spurious linkages can be identified and be removed.

The Zwicky Transient Facility (ZTF), an optical time-domain survey scanning the entire northern hemisphere of sky at a rate of more than 3700 $\deg^2$ $\text{hr}^{-1}$, can be seen as a precursor to the LSST \citep{Graham2019, Bellm2019, Masci2019}. ZTF uses the ZTF Moving Object Discovery Engine \citep[ZMODE;][]{Masci2019} algorithm. Instead of linking tracklets directly into tracks, ZMODE first attempts to build a ``stringlet". A stringlet forms an intermediate step between tracklets and tracks which allows for the linking of pairs of detections across nights before tracks are built. This approach was designed to accommodate ZTF's cadence during its main survey, where the cadence is frequently too sparse to form short intra-night tracklets. 

Recent work by \cite{Holman2018}, has shown promising results by shifting the reference frame for linking detections to the heliocenter. By assuming a heliocentric distance and its rate of change, cleverly fitting tracklets for the remaining unknown parameters in inertial space, then propagating the resulting ``arrows" to a common epoch, arrows corresponding to the same minor planets will form clusters. These clusters can then be extracted and subsequently validated by orbit fitting. As a testament to the effectiveness of HelioLinC, some 200,000 new minor planet orbits were recovered from the Minor Planet Center's Isolated Tracklet File (ITF) \citep{Holman2018}.

Common in all of these approaches is the requirement to build tracklets, which in turn requires a telescope to perform multiple revisits to the same field in a night, then more revisits a few nights later, and so on. For a survey that cannot cover the entire visible sky twice per night, this leads to up to a factor of two reduction of the nightly surveyed area. For a survey such as the LSST, which aims to balance four different science drivers, requiring such a cadence decreases the overall ease by which the other science drivers can be accommodated. It is therefore prudent to investigate whether linking algorithms that are cadence independent can be constructed and whether such algorithms can perform as good or better than the current methods. An algorithm that does not demand a high revisit cadence could increase the efficiency of future surveys, as well as help multi-science missions such as the LSST.
\\

In this paper, we present one such cadence- and observer-independent linking algorithm: ``Tracklet-less Heliocentric Orbit Recovery" (THOR). Rather than shifting the origin to the heliocenter like in \cite{Holman2018}, we choose to shift the linking frame of reference to a series of dynamically selected heliocentric ``test orbits''. The main insight is that transforming detections into the frame of a test orbit linearizes the motion of all objects in a relatively thick bundle of orbits near the test orbit (in phase space), which can then be picked out with line-detection algorithms such as the Hough transform. This provides a path to scanning an otherwise voluminous 6D phase space with a finite number of test orbits and at feasible computational cost.

We describe the THOR algorithm in Section \ref{sec:algorithm}. We test its performance on simulated and real survey observations in Sections \ref{sec:simulations} and \ref{sec:ztf}, respectively. Lastly, we discuss planned extensions and future work in Section \ref{sec:discussion}.

\section{Algorithm} \label{sec:algorithm}

Algorithms which rely on tracklets to discover moving objects generally search for the subset of orbits that are consistent with the observed tracklets and their constituent observations. THOR takes a different approach -- it aims to identify subsets of observations that are consistent with being close to one or more ``test orbits'' drawn from the 6D orbital phase space. For each such test orbit, THOR searches for detections which lie near or adjacent to it as it is propagated backward or forward in time using clustering and line-finding algorithms (see Section \ref{sec:hough}).

This approach is computationally feasible because a single test orbit can be used to recover all objects that exists in a finite (and relatively large) ``bundle'' around it in phase space. This makes it possible to select and explore a sizable volume of phase space with a finite number of test orbits. It also makes it possible to focus on regions of phase space where we expect to find objects or that we know are populated, which we take advantage of to computationally efficiently search for specific asteroid populations in ZTF (Section~\ref{sec:ztf}).

There are several advantages to this approach to minor planets searches. First, the propagation of the test orbit is only limited by the robustness of the used orbit propagator. For example, propagating the test orbit using an N-body integrator allows the full gravitational perturbation by the Sun, planets, and massive minor planets to be properly taken into account. This is particularly useful for NEOs or any minor planets that might have non-trivial orbital encounters. Second, a fully defined orbit can be propagated to epochs beyond the scope of any one survey, permitting the use of THOR on archival and combined datasets\footnote{This also means that THOR is observer-independent from the standpoint of linking detections from multiple observers.}. Third, by searching for observations near an assumed test orbit, the cadence of observations becomes irrelevant. The orbit can simply be propagated to any epoch where an observation is possible regardless of the temporal spacing between observations themselves. This removes the requirement for tracklets to be observed, and enables the linking of multiple single-night detections instead of the linking of multiple pairs of nightly detections.

The THOR approach does come with challenges. The computational cost, compared to HelioLinC-type algorithms, is higher due to the need to perform reference frame transformations to a (potentially large) number of test orbits rather than a single transformation to the heliocenter. However, as we show in Section~\ref{sec:ztf}, this cost can be readily met in practice.
A more challenging issue is that of the selection of test orbits to optimally cover the phase space of interest (the phase space where small bodies are likely to be found). While later in this section we show that a relatively ad-hoc approach to orbit selection (essentially, importance sampling from the known population) is a very effective way to scan for all objects in a known population, it is also clear this approach will leave the algorithm blind to unknown populations or objects on unusual orbits\footnote{Though, occasionally, such objects are still serendipitously found, as witnessed by the recoveries of NEOs in our test dataset.}. We leave the problem of developing an algorithm to select the optimal number of orbits to {\em exhaustively} search the entire phase space for future work.

\subsection{THOR}
We next describe the five main components of the THOR algorithm that enable the discovery of moving objects using a test orbit. We explain how test orbits are selected from 6D orbital phase space at the end of this section. 

\subsubsection{Select a Test Orbit}\label{sec:testorbit}

\begin{figure}[hbt!]
  \centerline{\includegraphics[width=0.45\textwidth,keepaspectratio]{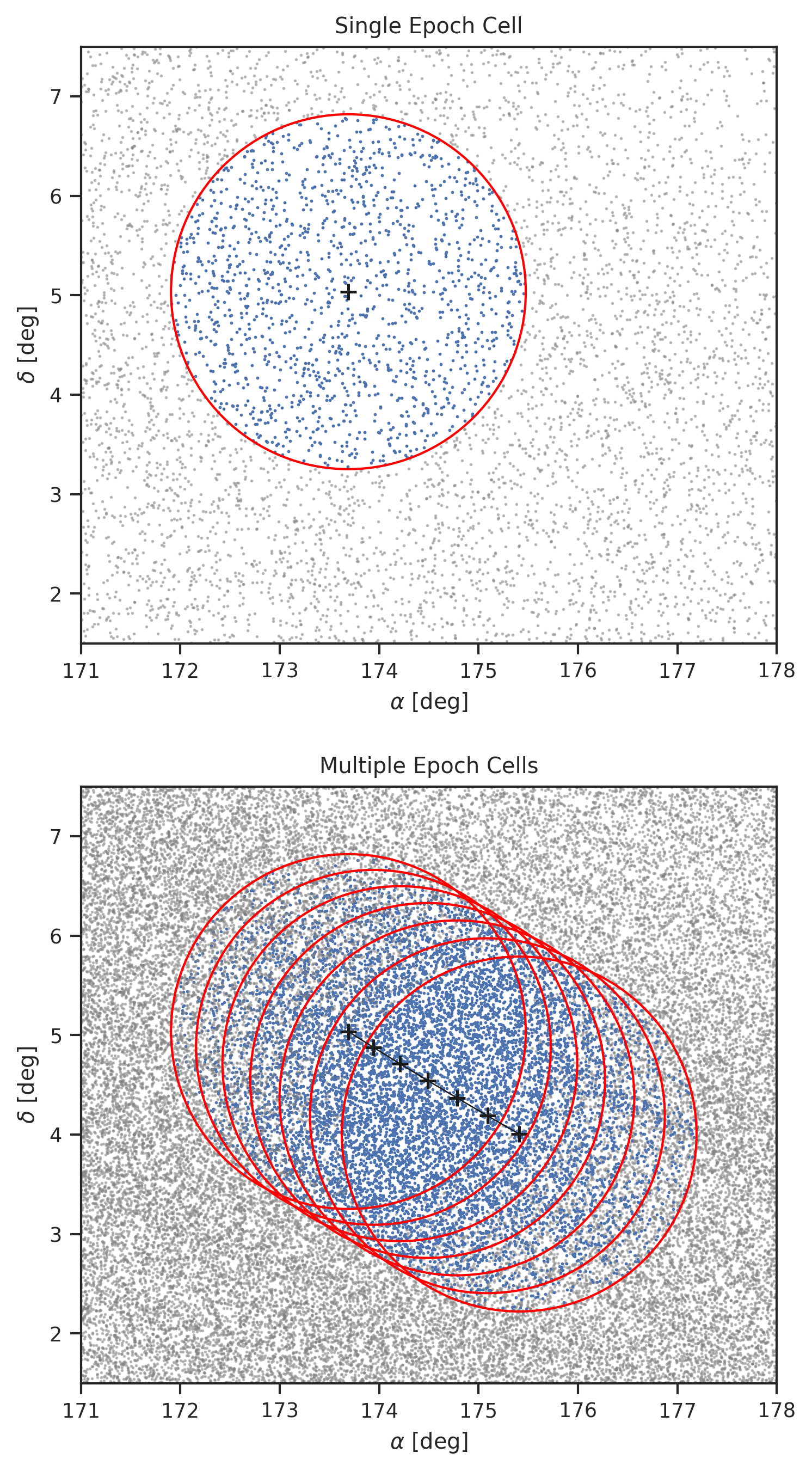}}
  \caption{In the top panel, simulated detections of real orbits on the first night of a simulated survey are plotted in grey. The survey consists of 16 ten $\deg^2$ fields visited once every other night over a 14-night window. The location of the test orbit on the first night is shown as a black plus sign. The red circle outlines the cell of gathered detections which are plotted in blue. In the bottom panel, the test orbit is propagated to all possible times in the survey (the remaining six possible exposures) with a cell of observations gathered at each predicted location and epoch. The simulated detections of the subsequent six visits are plotted in grey in addition to those from the first night. The gathered detections are plotted in blue as in the top panel. The black line tracks the sky-plane motion of the test orbit, with its location on each line plotted as black plus signs. This figure was generated using \href{https://nbviewer.jupyter.org/github/moeyensj/thor_notebooks/blob/master/paper1/plots_simulations.ipynb}{plots\_simulations.ipynb}.}
  \label{fig:algorithm12}
\end{figure}

A test orbit with heliocentric position vector, $\vec{r}$, and heliocentric velocity vector, $\vec{v}$, at some epoch, $t$, is placed in a survey that contains the detections of moving objects. In the top panel of Figure \ref{fig:algorithm12}, we show an example test orbit placed in the simulated survey described in Section \ref{sec:simulations}.  

\subsubsection{Propagate Test Orbit and Gather Detections}\label{sec:propagate}

The test orbit is propagated to all possible epochs in the survey. For each unique epoch all detections within some area, $A$, measured on the sky-plane are gathered. We term the area, $A$, a ``cell" of observations where the test orbit could be representative of the underlying orbital distribution. 

Figure \ref{fig:algorithm12} shows the topocentric sky-plane motion of a test orbit when propagated through a simulated survey. The black line tracks the sky-plane trajectory of the test orbit. The red circles outline cells of observations appearing at compatible times in the survey.

\subsubsection{Heliocentric Transformation and Projection into the Test Orbit's Frame of Reference}\label{sec:transformproject}

All cells of detections gathered, they can be transformed into the heliocentric frame, rotated and then projected onto a gnomonic tangent plane centered about the motion of the test orbit. We refer the reader to Appendix \ref{sec:appendix} for details on this procedure.

\begin{figure*}[htb!]
  \centerline{\includegraphics[width=0.9\textwidth,keepaspectratio]{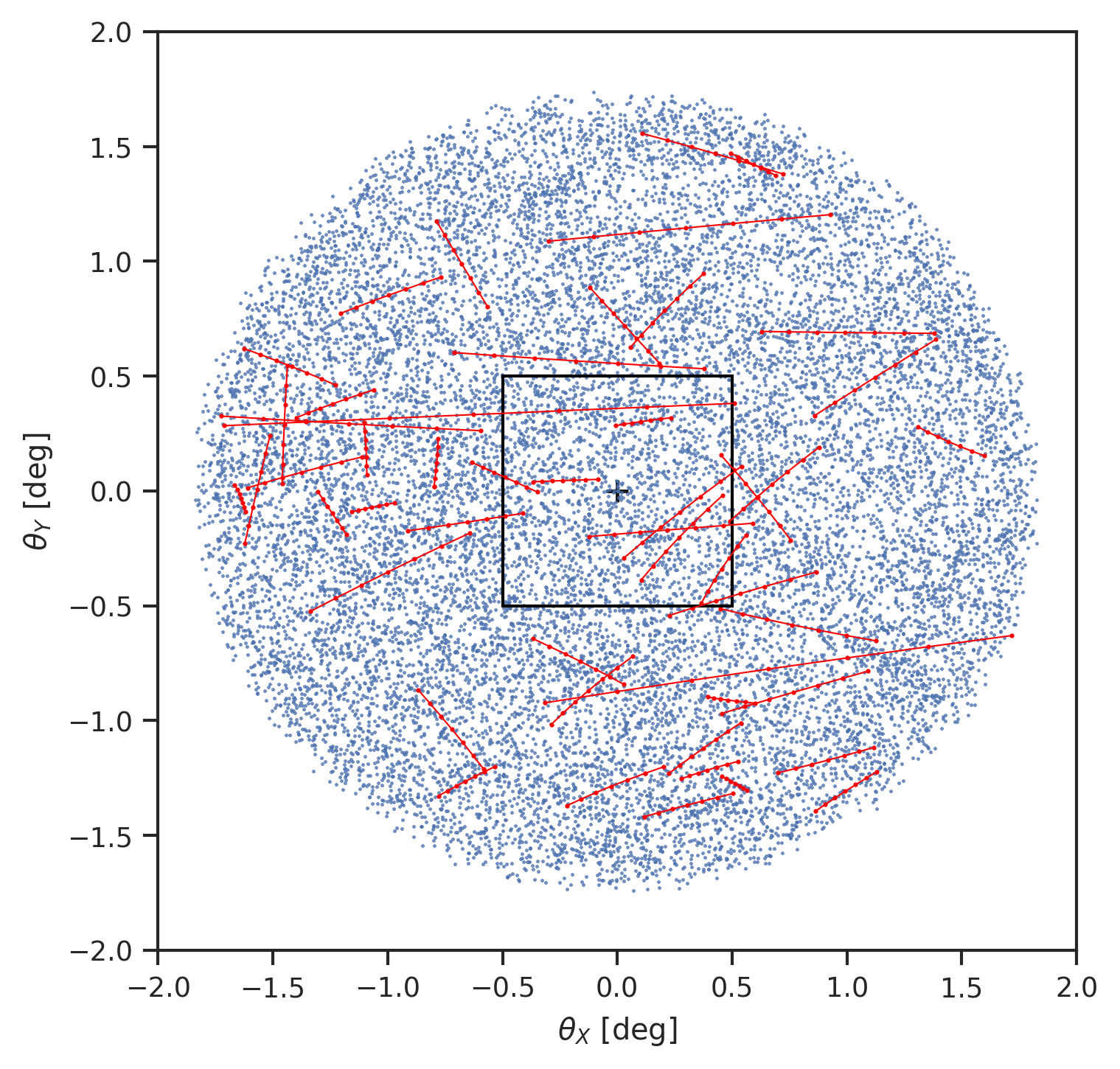}}
  \caption{Plotted in blue are the cells of detections from Figure \ref{fig:algorithm12} after they have been transformed to the heliocentric frame and then rotated and projected into the frame of the test orbit's motion. Once transformed and projected, the cells can be stacked. Note that the co-rotating reference frame is fundamentally three dimensional. The time dimension (z-axis) has been collapsed onto the gnomonic plane ($\theta_X$-$\theta_Y$ plane). In its own frame of reference the test orbit (black plus sign) lies at (0, 0). Objects with orbits similar to the test orbit will appear as lines or curves, some of which are visible in this image. We plot the detections of 50 such objects in red, with their underlying true linkages plotted as red lines. There are a total of 1,397 objects with at least five detections in this image, these are considered findable. The black square outlines the subset of the data used to describe how clustering is performed in Figure \ref{fig:algorithm4}. This figure was generated using  \href{https://nbviewer.jupyter.org/github/moeyensj/thor_notebooks/blob/master/paper1/plots_simulations.ipynb}{plots\_simulations.ipynb}.}
  \label{fig:algorithm3}
\end{figure*}

Figure \ref{fig:algorithm3} shows the transformed, rotated and then projected detections in the co-rotating frame of the test orbit. Both $\theta_X$ and $\theta_Y$ are measured relative to the location of the test orbit at each unique epoch. Here $\theta_X$ lies along the plane of the test orbit, whereas $\theta_Y$ is perpendicular to the plane of the test orbit. Once the detections have been transformed to the frame of the orbit and projected onto the tangent plane, each cell of transformed observations can be stacked. Any object that has an orbit similar to that of the test orbit will appear as a cluster: a circular group in the $\theta_X$-$\theta_Y$ plane (or equivalently, a line in $\theta_X$-$\theta_Y$-t space). In the limit where the test orbit is identical to the orbit of an object in the frame centered on the test orbit's motion, that object will appear as a cluster with a radius proportional to the mean astrometric uncertainty of its constituent detections. Any objects that are on orbits near to the test orbit will appear as lines or curves in the co-rotating frame (there are a few such examples visible in Figure \ref{fig:algorithm3}).

\subsubsection{Hough Transform}\label{sec:hough}

The generalized Hough transform (GHT) is a template matching algorithm that can extract arbitrary features from multi-dimensional space \citep{BALLARD1981111}. Lines and curves can be extracted from the test orbit's co-rotating frame of reference using the equivalent of a 3D generalized Hough transform in $\theta_X$, $\theta_Y$, and time, $t$, space. To do so, we make a 2D velocity grid in $d\theta_X/dt$ and $d\theta_Y/dt$. By shifting the detections using a grid of assumed linear velocities relative to the test orbit's motion we effectively perform a 3D Hough transform as a 2D clustering problem. 

\begin{figure*}[!hbt]
  \centerline{\includegraphics[width=1\textwidth, keepaspectratio]{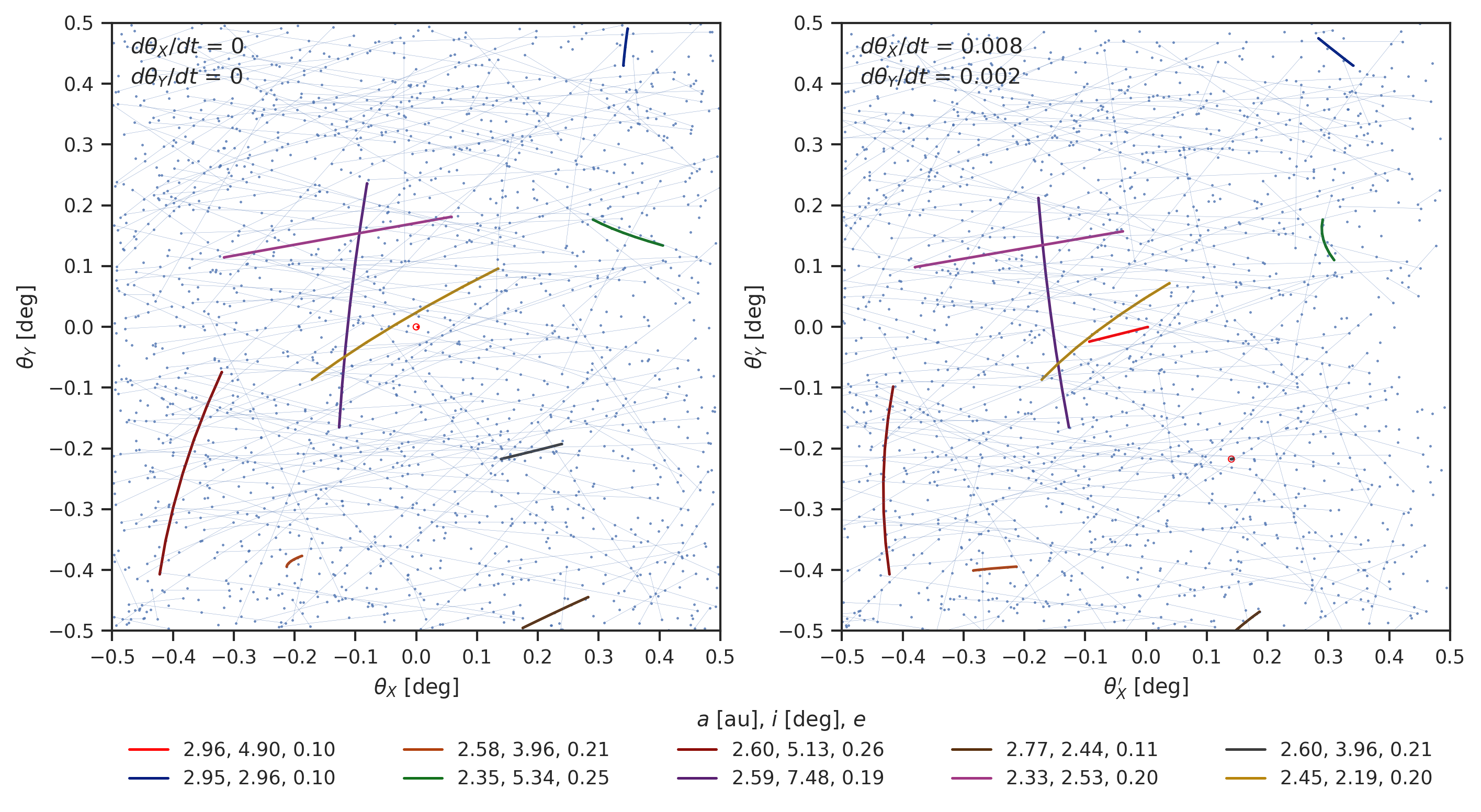}}
  \caption{By assuming some linear velocity in the co-rotating frame of the test orbit, objects with similar orbits can be recovered. In the left panel, we plot the same data as outlined by the black square in Figure \ref{fig:algorithm3}. The test orbit is the red cluster of points centered on (0, 0). The red circle surrounding the cluster is the effective cluster radius used (0.005 $\deg$). In blue are the detections of other objects with their true underlying linkages also shown. We plot nine of these other true linkages as colored lines to highlight how we can find other objects. In the right panel, we shift the detections by assuming some velocity in both $\theta_X$ and $\theta_Y$ relative to the test orbit. An object that originally appeared as a line in the test orbit's projected space now appears as a cluster. Note as well, that since our test orbit was a real object, in the non-zero velocity frame it appears as a line where in the zero velocity frame it appears as a cluster. This figure was generated using  \href{https://nbviewer.jupyter.org/github/moeyensj/thor_notebooks/blob/master/paper1/plots_simulations.ipynb}{plots\_simulations.ipynb}.}
  \label{fig:algorithm4}
\end{figure*}

In Figure \ref{fig:algorithm4} the subset of data outlined by the black square in Figure \ref{fig:algorithm3} is plotted. The left panel shows the co-rotating frame of the test orbit with the relative motion of nine other objects drawn as lines. The remaining detections with their underlying true linkages are shown in blue. Note a cluster shown in red at (0, 0). Surrounding the cluster is a circle indicating the chosen clustering radius. In the right panel, the same detections after they have been shifted assuming a linear velocity in both $\theta_X$ and $\theta_Y$ are plotted. Note how one of the nine other objects now forms a cluster; that cluster is also surrounded by a circle indicating the clustering radius (this cluster was found using the actual clustering code). Because the selected test orbit was the orbit of a real object, in the non-zero velocity frame its detections now also form a line as shown by the red line towards the middle of the right panel. 

With a fine velocity grid it is likely that similar test velocities will result in the building of clusters with identical subsets of detections. This is also a function of the maximum allowable clustering radius. After all velocities have been tested, the clusters are scanned and any duplicates that have the same constituent detections as another cluster are removed. 

We consider any cluster with detections belonging to a single object a ``pure" cluster. Some clusters may be contaminated with observations from artefacts or some other, unassociated, object(s). We call these ``partial'' clusters. For example, a cluster with a total of ten detections where eight of those detections belong to one object and two detections are either false positives or belong to another object or objects, would have a contamination percentage of 20\%. Partial clusters could still permit for the discovery of objects as outlier rejection can be used to filter out the imposter detections. Any cluster that is neither pure nor partial, is considered a ``mixed'' cluster. In the next section we describe how the process of orbit determination is used to test the validity of the clusters, and how outlier rejection is used to clean partial clusters.

\subsubsection{Orbit Determination}\label{sec:od}

Orbit determination in THOR is a three step process with the aim of testing the validity of the clusters produced by the Hough transform described in Section \ref{sec:hough}.

Initial orbit determination (IOD) serves as a first-level filter against erroneous linkages, intended to efficiently reduce the number of clusters to be considered for full differential orbit determination (OD). We implemented the method of Gauss following the formalism introduced by \cite{Milani2008}. It seeks to fit a Keplerian orbit to three observations on the sky by solving for the orbit at the time of the second observation. This involves solving an eighth order polynomial which can result in up to three solutions per linkage. A challenge for this approach is choosing which three observations to use: as our implementation is not computationally expensive, we elected to fit every combination of three observations for each cluster (i.e., a cluster of five constituent observations would mean testing ten combinations) until a combination is selected such that the $\chi^2_\nu$ with respect to the observations\footnote{We note that one must be careful when interpreting the distribution of $\chi^2$ values when only three observations are used to fit the preliminary orbit, but residuals are calculated against {\em all} $N$ observations in the cluster. The resulting distribution will not follow the $\chi^2$ distribution with $2 N-6$ degrees of freedom, but generally be wider.} meets a user-defined threshold. If all combinations are exhausted before the threshold is met, and no outliers can be removed then the candidate orbit is discarded.

To clean and filter ``partial'' linkages or clusters, we added outlier rejection to the IOD component of THOR's orbit determination pipeline. The user sets a contamination threshold which defines the maximum percent of observations in a linkage that can be flagged as outliers. If all combinations of three observations have been tested via the method of Gauss, the linkage has not yet met the acceptance threshold, and the linkage has sufficient observations to allow for outlier testing, then the observation with the highest $\chi^2$ residual is removed. If the orbit's overall $\chi^2_\nu$ meets the user-defined threshold, the orbit is accepted. If the orbit does not meet the threshold, the next highest residual observation is also removed, and so on, until the maximum number of observations have been removed as outliers or the threshold has been met.

The dynamic range of $\chi^2_\nu$ values obtained for clusters may span many orders of magnitude. This is also true even for pure clusters, where the approximate nature of the method of Gauss might yield residuals as high as several arcseconds. To further improve these solutions a re-fit could be attempted using a more robust technique such as the method of Herget \citep{Herget1965}. However, we found it sufficiently computationally efficient to simply pass all orbits satisfying a very relaxed IOD cut  to the differential corrector. The corrector then acts as a much stronger and accurate filter.

After IOD, the preliminary orbits are differentially corrected (OD) using the method of weighted linear least-squares in combination with numerical differencing to determine the Jacobian (the matrix of partials of the observations with respect to changes in the orbit state). We implemented the differential correction algorithm described in \cite{Vallado2013} with a few notable differences: we added outlier rejection in a similar fashion to the IOD code, a $\chi^2_\nu$ acceptance threshold, and adaptive finite differences.

The goal of differential correction is to improve a preliminary orbit solution by minimizing the residuals across all its constituent observations. The improved orbit is propagated to all observation times and residuals with respect to the observations are calculated. If such differentially corrected orbit meets a user-defined $\chi^2_\nu$ threshold, it is accepted. If the differentially corrected orbit does not meet the user-defined threshold, iteration continues until the maximum number of iterations is reached. If at this point the orbit does not meet the user-defined $\chi^2_\nu$ threshold, the observation with the highest $\chi^2$ value is removed and iterative improvement is continued using the new set of observations. The iterative outlier removal continues until the maximum allowable number of observations have been removed as determined by the user-settable contamination threshold. This process is an effective filter against erroneous (``mixed'') orbits and is an effective method to clean contaminated (``partial'') orbits.

Orbits that pass the differential correction step will likely contain only a subset of the observations that belong to those orbits. Some orbits also will share observations with other orbits (e.g. orbits that represent different sub-arcs of one longer arc). The last step in THOR's orbit determination process aims to merge orbits that represent the same object, and extend orbits that could have more observations associated with it. This is done through an iterative combination of OD and observation attribution. We perform attribution by propagating the differentially corrected orbits to all epochs of the observations contained in the gnonomic tangent plane of the test orbit. Any observations that lie within a user-defined angular distance of the predicted location are identified as possible attributions. If at this point any observations are identified as belonging to multiple orbits, these orbits are merged to form one larger orbit (a ``child'' orbit) and re-fit. If such re-fitted child orbit passes the OD-filter (meets the $\chi^2_\nu$ threshold), its parent orbits are removed from the active pool of possible orbits. Any orbits that pass OD but could not be improved further are unlikely to have more observations attributed to it and are output as high quality candidate orbits. This iterative combination of attribution and re-fitting is done until each orbit has only unique constituent observations and the orbits are not further improved by OD.
\\

In summary, the orbit determination stage filters and transforms the candidate clusters produced by the Hough transform step into a catalog of high quality differentially corrected orbits. These successfully linked orbits and their observations are removed from the active pool of observations, and steps \ref{sec:testorbit} - \ref{sec:od} are repeated with new test orbits to attempt to link the remaining observations.

\subsection{Phase-Space Gridding}\label{sec:test_orbits}

\begin{figure}[hbt!]
  \centerline{\includegraphics[width=0.53\textwidth,keepaspectratio]{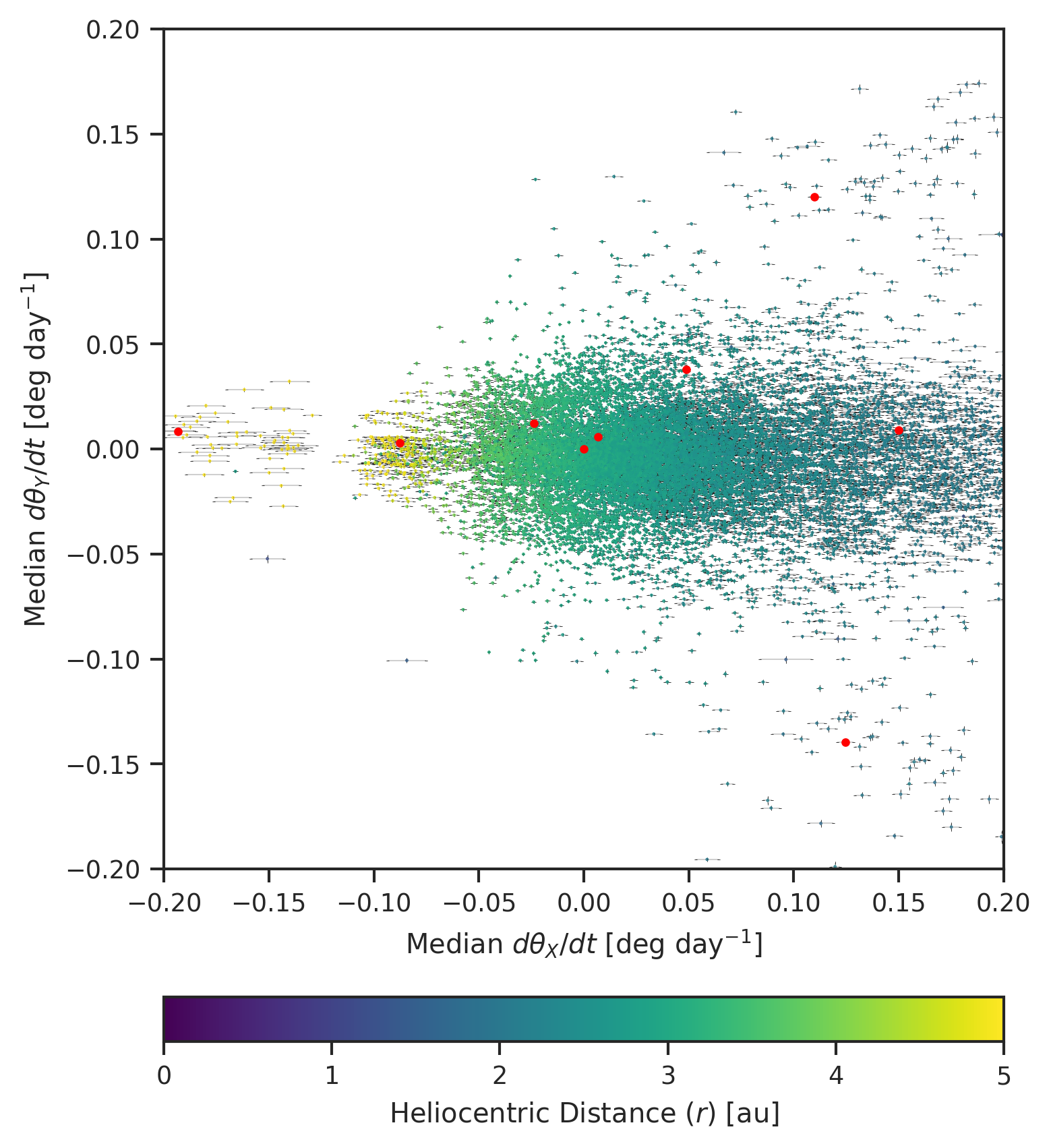}}
  \caption{Using one of the test orbits from the simulated survey but now setting the cell size to 500 $\deg^2$, we plot the median velocities in $\theta_X$ and $\theta_Y$ relative to the test orbit for each findable object in the co-rotating frame of the test orbit. The error bars are the respective standard deviations of $d\theta_X/dt$ and $d\theta_Y/dt$, which serve as a proxy for the curvature in the projected space. The points are colored by the median heliocentric distance. The red points show the locations of the test orbits selected for the survey using the orbit selection criteria described in Section \ref{sec:test_orbits}. As before, the test orbit on which this image is centered has a velocity of (0, 0) $\deg \text{ day}^{-1}$. The default velocity grid is set to 300 equally spaced bins between -0.1 and 0.1 $\deg \text{ day}^{-1}$ in both coordinates. The range of the velocity grid allows different heliocentric distances to be probed, which in effect allows one test orbit to recover other objects as long as their motion relative to the test orbit appears linear and the rate of motion lies within the searched grid. This figure was generated using  \href{https://nbviewer.jupyter.org/github/moeyensj/thor_notebooks/blob/master/paper1/plots_simulations.ipynb}{plots\_simulations.ipynb}.}
  \label{fig:orbit_frame_velocity_500}
\end{figure}

A key element of THOR is the selection of test orbits and how to best populate the relevant phase space to maximize completeness for different populations of minor planets. The bulk of the Solar System small bodies that LSST will discover are Main Belt asteroids \citep{Jones2018}. We initially focus on populating the phase space to discover populations in the Main Belt and the slower-moving outer Solar System populations. Future work is planned to target the faster-moving NEO population, and to further optimize test orbit selection.

Guided by tests on real ZTF data, the current implementation of THOR uses nine average orbits from the known population of minor planets per patch of sky. The patches of sky are currently a simple rectangular sky-plane subdivision 15 by 15 $\deg$ in ($\alpha$, $\delta$) with no overlap. This notional subdivision could easily be replaced by something more robust such as HEALPix \citep{Gorski2005}. The cell size was set to 1000 $\deg^2$ to maximize the number of detections gathered by each test orbit at each epoch. This size permits a test orbit to be selected along any edge of a patch and still gather all detections in that patch of sky during the first epoch of observation. The patches in which an average orbit is calculated do not have any effect on the detections that propagating a test orbit can gather, they are simply used to find a set of test orbits.

The test orbits are selected in bins of semi-major axis ($a$), with bin edges roughly corresponding to the location of the most prominent mean-motion resonances (the Kirkwood gaps). To target the Main Belt we set the inner most edge at $a = \text{1.7 au}$ with the outermost edge at $a = \text{5.0 au}$. The remaining bin edges were set to $a = \text{2.06, 2.5, 2.82, 2.95, 3.27 au}$, which correspond roughly to the Saturn $v_6$ secular resonance, and the 3:1, 5:2, 7:3 and 2:1 Jovian mean motion resonances, respectively \citep{Minton2009}. To target the outer Solar System, we added a single bin with edges $a = \text{5.0 au}$ and $a = \text{50.0 au}$.

For each bin, an object with average eccentricity ($e$) and inclination ($i$) is selected from the Minor Planet Center's  (MPC) catalog of known objects. No test orbits with eccentricity greater than $0.5$ were selected. This relatively simple orbit selection is not sufficient to target the complete Main Belt population. To target the missed populations from earlier trials we sub-divided the first semi-major axis bin further, in particular, the bin located at the approximate location of the Hungaria family of asteroids ($e < 0.18$, $1.78 < a < 2.00 \text{ au}$, $16 < i < 34 \text{ deg}$). The Hungarias are notable for their wider spread of orbits in orbital phase space, especially in inclination \citep{WARNER2009172}. Three test orbits were selected in different bins of eccentricity ($e$) with bin edges (including the exterior edges) $e = \text{0.0, 0.1, 0.2, 0.4}$.

In Figure \ref{fig:orbit_frame_velocity_500} we plot the median linear velocities in $\theta_X$ and $\theta_Y$ of all the findable objects relative to one of the test orbits selected for the simulated survey. The simulated survey is described in the following section. The plotted error bars are the standard deviations of the velocities and are a proxy for the curvature in the 3D space of the gnomonic plane and time. The test orbits that were selected for use on the simulated survey are plotted in red. We used many different iterations of Figure \ref{fig:orbit_frame_velocity_500} to inform decisions regarding the test orbit selection code. The test orbit selection criteria and code were designed in such a way as to maximize coverage in this space. 

Note that by choosing different upper and lower velocity bounds, different heliocentric distances are probed by each test orbit. For example, the velocity grid can be set to sweep out a wider heliocentric distance at the cost of greater computation time. Equivalently, an additional test orbit could be used with a smaller grid to the same effect and at a possibly lower computational cost. These are trade-offs which may depend on the survey and its data properties, in particular, a survey's observational depth influences the number of observations, and its hardware and software characteristics influence the occurrence of false positives. In general, the propagation and gathering of detections takes of order seconds on a single thread, whereas the current Hough transform implementation takes of order minutes to hours with moderate parameters on a few dozen threads. The Hough transform component of THOR is where future optimization will be focused in addition to improving test orbit selection.
\vspace{0.2in}

\section{Validation with simulated data}\label{sec:simulations}

\subsection{Simulation Setup}

\begin{figure}[hbt!]
  \centerline{\includegraphics[width=0.45\textwidth,keepaspectratio]{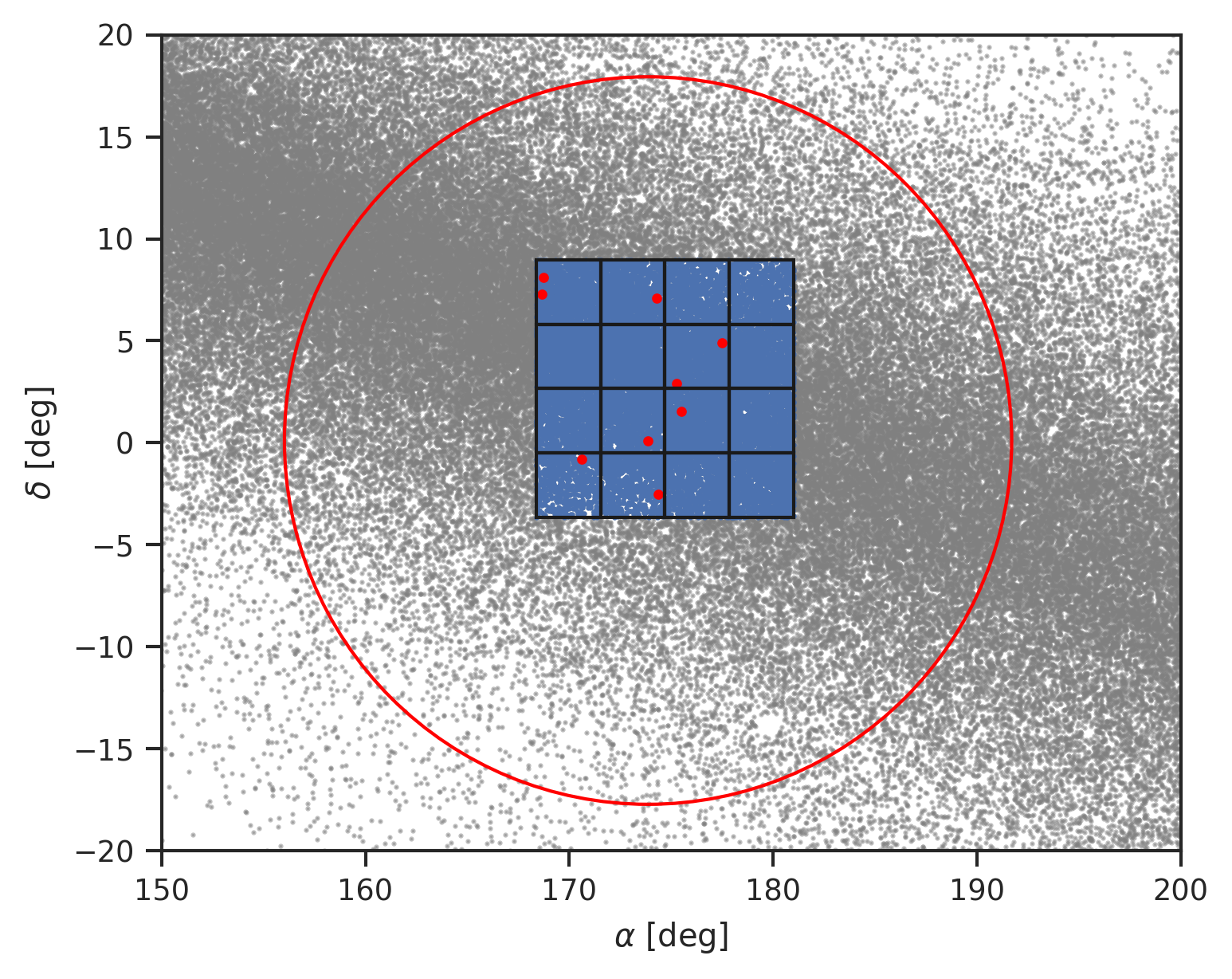}}
  \caption{Plotted in blue are the simulated detections of real orbits generated using the MPC's orbit catalog for the first night of the simulation. The ephemerides for the underlying orbit catalog from which the blue detections are drawn are shown in grey. The red points indicate the locations of the selected test orbits on the first night. The red circle traces the 1000 $\deg$ cell size for one of the orbits. The 16 square fields (black squares) each ten $\deg^2$ in size are visited once every other night over a period of 14 nights. The footprint was roughly centered on the ecliptic to maximize minor planet density. No false positive detections are plotted even though they were included in the simulated survey. This figure was generated using  \href{https://nbviewer.jupyter.org/github/moeyensj/thor_notebooks/blob/master/paper1/plots_simulations.ipynb}{plots\_simulations.ipynb}.}
  \label{fig:simulations_footprint}
\end{figure}

To test the efficacy of our algorithm we created a simulated mini-survey composed of 16 square fields ten $\deg^2$ in size. Each field is visited once every other night over a 14 day period yielding seven unique visits per field over the course of the simulated survey. We downloaded the Minor Planet Center's catalog\footnote{\url{https://minorplanetcenter.net/iau/MPCORB.html}} of known orbits and used open-source orbit propagation and determination software, \textit{oorb}, to generate ephemerides for each known object at the different survey times \citep{oorb}. In practice, THOR will run on the detections resulting from difference imaging, however, simulated ephemerides are sufficient to test the performance of linking detections.

The survey footprint was roughly centered on the ecliptic to maximize the density of minor planet detections (see Figure \ref{fig:simulations_footprint}). We added $100 \pm 10$ randomly distributed false positive detections per squared degree to simulate the effect of image differencing artefacts. The final survey consisted of 139,120 (55.4\%) simulated observations belonging to real orbits and 112,078 (44.6\%) false positive observations. The simulated observations were randomized with 100 mas scatter to reflect the expected astrometric errors for moving objects as observed by current generation surveys.

We consider any minor planet with at least five detections throughout the survey to be findable, yielding a total of 18,332 such objects. Any object that is recovered in a pure orbit is considered to be found. For comparison, the Vera C. Rubin Observatory's current implementation of MOPS would require three tracklets (six detections in three intra-night pairs) for an object to be findable -- as a tracklet based algorithm it would recover none of the objects in the simulated survey.

The contamination threshold for partial clusters was set to 20\% with a minimum cluster size of five detections (see Section \ref{sec:hough}). The default grid in both $d\theta_X/dt$ and $d\theta_Y/dt$ for the Hough transform was set to 300 equally binned velocities between -0.1 and 0.1 $\deg \text{ day}^{-1}$ relative to the motion of the test orbit, yielding a total of 90,000 velocities tested per test orbit. The minimum cluster size was set to five detections (equal to the findability criterion), with a cluster radius of 0.005 $\deg$. To extract clusters for every velocity tested we use \textit{scikit-learn}'s implementation of the Density-Based Spatial Clustering of Applications with Noise  (\textit{DBSCAN}) algorithm \citep{DBSCAN, scikit-learn}. DBSCAN is a non-parametric clustering algorithm that finds clusters in higher-dimensional space of approximately equivalent density given a maximum distance parameter and marks un-clustered data as outliers. We use open-source orbit determination software, \textit{oorb}, to propagate the test orbits to all possible times in the survey \citep{oorb}.

For orbit determination, the $\chi^2_\nu$ and contamination thresholds were set to ($10^5$, 20\%), ($10$, 20\%), and ($10$, 0\%) for IOD, OD, and the iterative combination of differential correction and attribution, respectively. The maximum attribution distance was set to one arcsecond. The minimum number of observations and minimum arc length for an orbit to be accepted was set to five observations and one day, respectively.

\subsection{THOR Performance}

Using nine test orbits, selected as described in Section \ref{sec:test_orbits}, THOR recovered 16,729 candidate orbits representing 16,728 (91.3\%) of the findable objects in the simulated survey. Only one erroneous orbit was produced.

\begin{figure*}[htb!]
  \centerline{\includegraphics[width=0.60\textwidth,keepaspectratio]{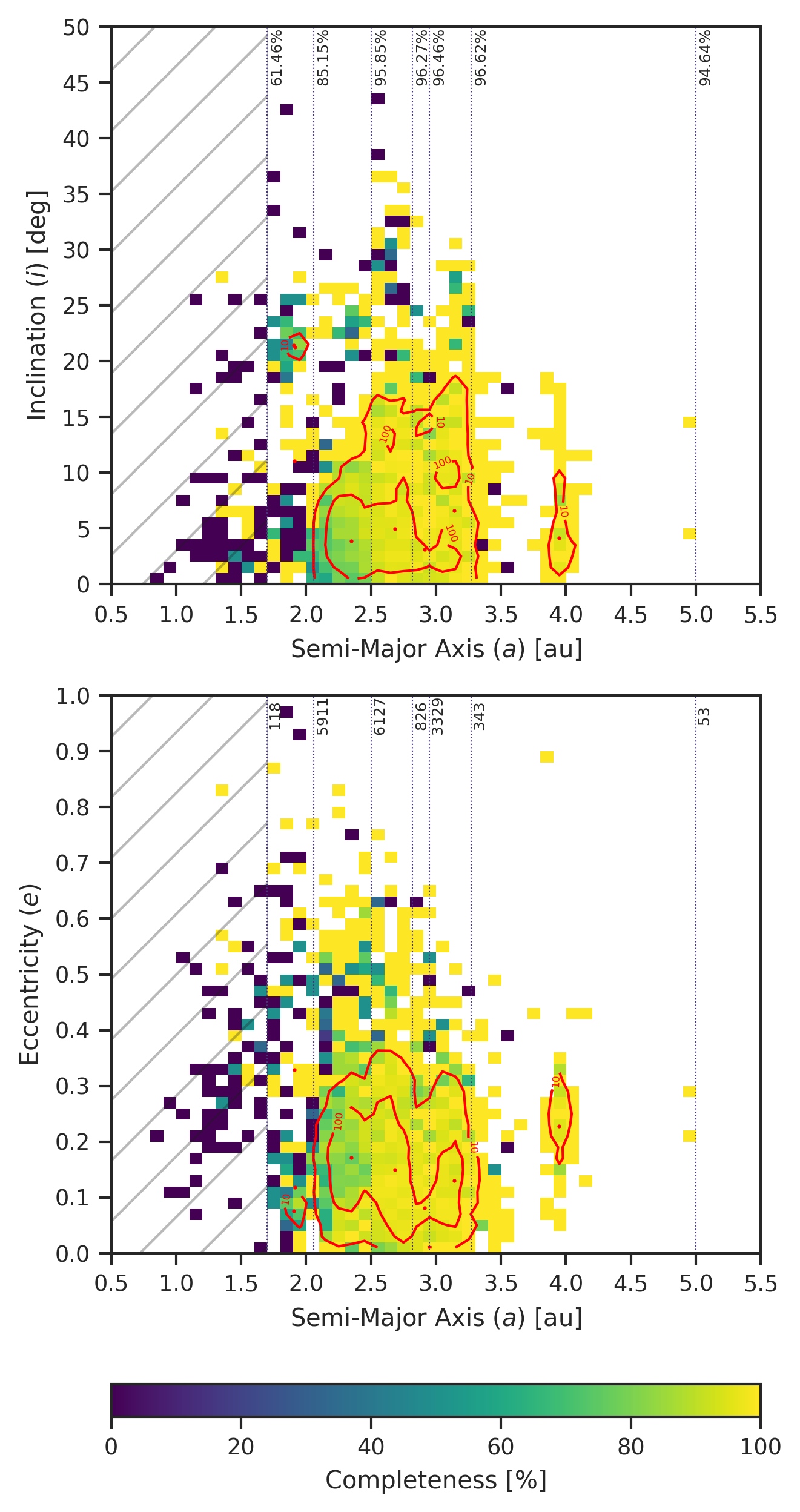}}
  \caption{In the top panel, percentage of objects found in pure orbits is visualized in bins of semi-major axis ($a$) and inclination ($i$). Number density contours are drawn as red lines to show the number of objects findable (five or more simulated detections through out the survey). The nine test orbits used are plotted as red points. The vertical dashed lines indicate the five semi-major axis bin edges, with the overall percentage completeness per bin written at the top. The hatched area indicates the region of semi-major axis space where no test orbits were chosen. In the bottom panel, percent completeness in bins of semi-major axis ($a$) and eccentricity ($e$) is plotted, contours and test orbits are plotted in the same style as in the top panel. Instead of percent completeness in each of the five bins we now explicitly state the number of objects found. This figure was generated using  \href{https://nbviewer.jupyter.org/github/moeyensj/thor_notebooks/blob/master/paper1/plots_simulations.ipynb}{plots\_simulations.ipynb}.}
  \label{fig:simulations_a_ie_completeness}
\end{figure*}

The top panel of Figure \ref{fig:simulations_a_ie_completeness} shows the recovered orbit completeness in bins of semi-major axis ($a$) and inclination ($i$). The red contours indicate the number of objects that should be findable, while the red points show the selected test orbits. The vertical dashed lines trace the semi-major axis bin edges. The percentages state the overall completeness in each bin of semi-major axis (note that the first bin has three test orbits). In the bottom panel, we repeat the same plot configuration as in the top panel, however, we plot eccentricity ($e$) instead of inclination. Instead of percentage completeness in the five semi-major axis bins we state the numbers of unique objects found to give the percentages context. As shown in both panels, the orbits selected in the bins beyond 2.5 au recover the vast majority (96.1\%) of the findable objects within that region. Performance in the semi-major axis bin between 2.06 au and 2.5 au (85.2\%) suggests that either more test orbits are needed or that the test orbit we selected is not adequately representative of the underlying orbital distribution. The same is true for the three orbits selected in the semi-major axis bin between 1.7 au and 2.06 au (61.5\%). Unsurprisingly, the two Hungaria orbits did not completely recover their constituent population because of the larger spread of Hungaria orbits in orbital element phase space. The high eccentricity orbit selected did, however, enable other high eccentricity objects to be recovered in addition to many objects in bins with larger semi-major axis. The test orbits were run in series starting in order of increasing semi-major axis. The second orbit that ran was the one selected in the inner Main Belt and over 10,000 of the findable objects in the survey were successfully found by using this test orbit. Overall completeness for the Main Belt and beyond ($a > 1.7$ au) is 91.6\%. While we did not focus test orbit selection to tackle the NEO population and populations interior to the Main Belt ($a < 1.7$ au), serendipitous completeness for these objects is 18.8\%.

\subsection{Computational Performance}

The complete simulations run, starting from a catalog of simulated observations to producing a catalog of 16,729 orbits using nine test orbits took seven hours on a personal workstation using 60 threads on 30 cores. The workstation has the following specifications: a 32-core/64-thread 2990WX AMD Threadripper, 128 GB DDR4 RAM, and 2 TB of SSD drive space. The run required a moderate amount of memory especially during the attribution of orbits, approximately 1.5 GB per thread. Given the heavy reliance of the current implementation on Pandas data frames, it is likely the memory usage could be further optimized. Theoretically, the algorithm should require little more 2-3x the space occupied by the observations.

\subsection{Completeness and Purity Considerations}

\begin{figure}[hbt!]
  \centerline{\includegraphics[width=0.5\textwidth,keepaspectratio]{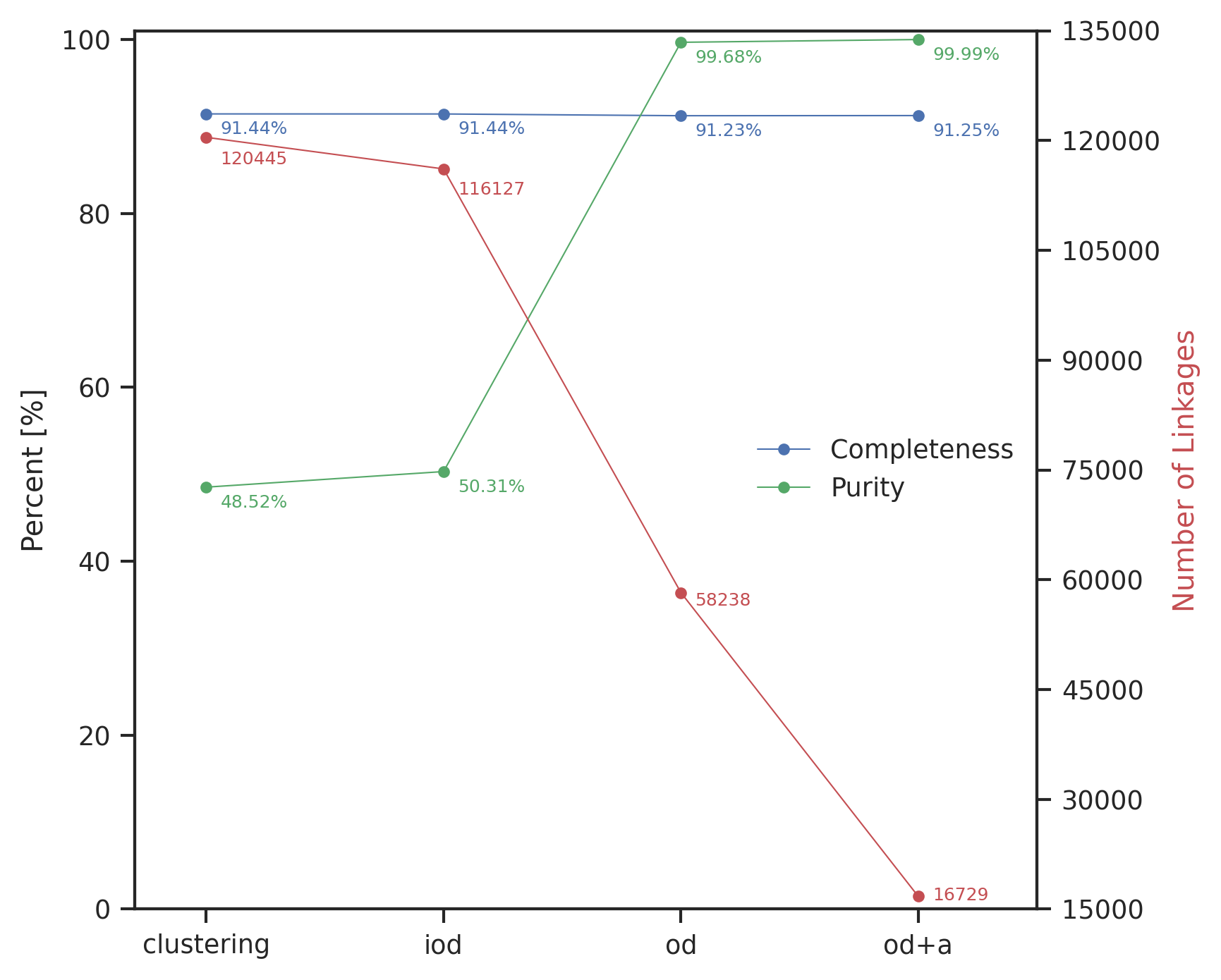}}
  \caption{Completeness and purity as functions of pipeline components are plotted in blue and green, respectively. The number of linkages is plotted in red. Each subsequent component of THOR enhances linkage purity while completeness remains largely unchanged. The number of linkages quickly approaches the true number of recovered objects. This figure was generated using  \href{https://nbviewer.jupyter.org/github/moeyensj/thor_notebooks/blob/master/paper1/plots_simulations.ipynb}{plots\_simulations.ipynb}.}
  \label{fig:simulations_component_summary}
\end{figure}

In Figure \ref{fig:simulations_component_summary}, we plot completeness (in blue) and purity (in green) as functions of the different pipeline components. Plotted in red is the number of linkages at each stage. From clustering to the final stage, we see only a 0.2\% drop in completeness, while the purity increases from 48.5\% to 99.9\%. Moreover, the number of clusters, initial orbits, and recovered orbits, quickly approaches the true number of objects recovered at each subsequent stage of the pipeline. The nearly constant completeness across the different pipeline components indicates that further completeness is likely to be found by improving our test orbit selection algorithm.

\section{Validation with Zwicky Transient Facility Data}\label{sec:ztf}

\begin{figure*}[htb!]
  \centerline{\includegraphics[width=0.95\textwidth,keepaspectratio]{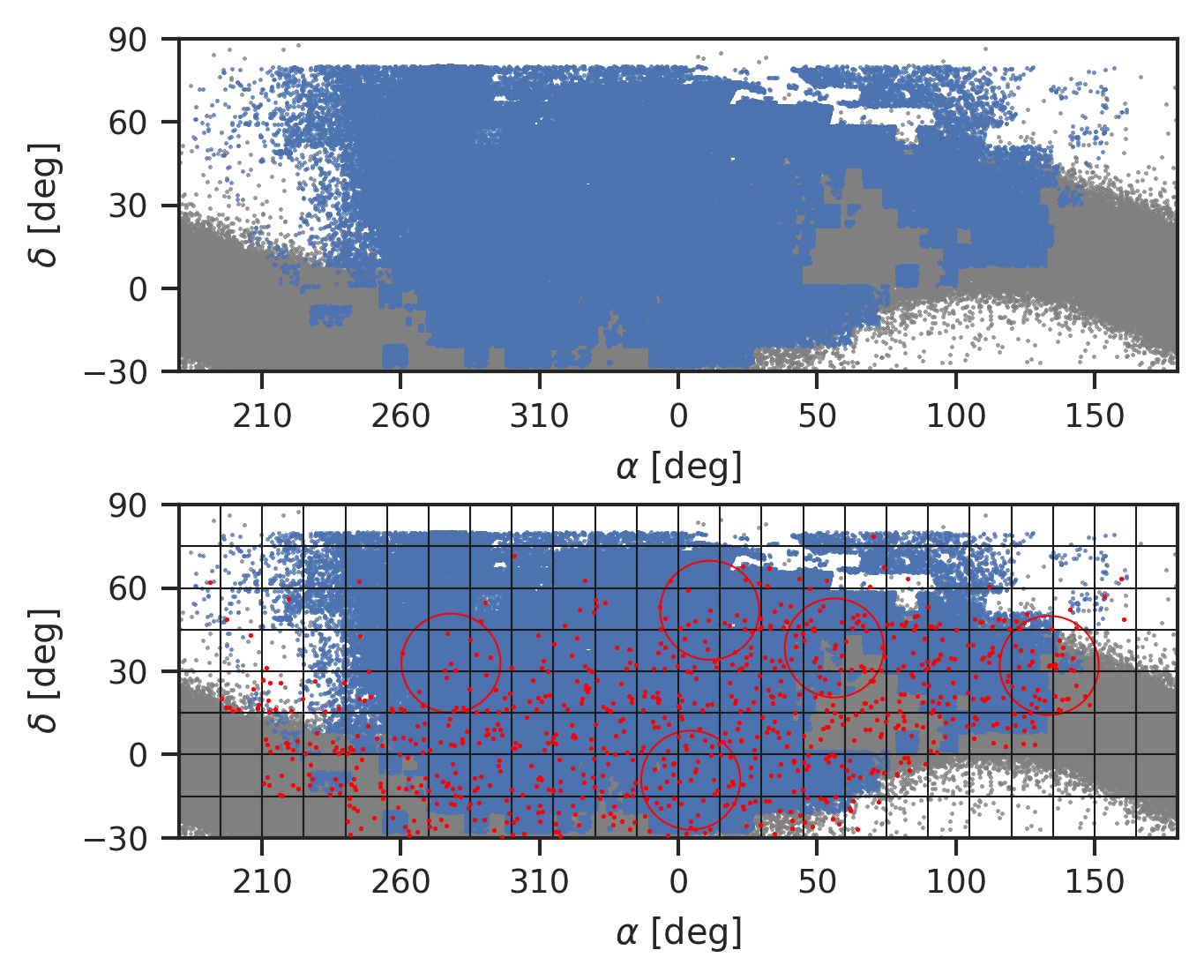}}
  \caption{In both panels, 15 nights of filtered ZTF alerts are plotted in blue. The catalog of known orbits with ephemerides generated for the first night of alerts by ZTF are plotted in grey. From the grey population test orbits are selected. In the bottom panel, we plot the same as the top panel but show the chosen sky-plane subdivision for test orbit selection (the 15x15 $\deg^2$ patches). For each patch, test orbits (in red) from the catalog of known objects were selected. If no ZTF observations occurred in any individual patch, no test orbits were selected. If no known object is predicted to be in any patch or bin of semi-major axis, no test orbit was selected. We randomly selected five of the test orbits and outline their 1000 $\deg^2$ cells of observations as red circles. This figure was generated using \href{https://nbviewer.jupyter.org/github/moeyensj/thor_notebooks/blob/master/paper1/plots_ztf.ipynb}{plots\_ztf.ipynb}.}
  \label{fig:ztf_footprint}
\end{figure*}

The Zwicky Transient Facility (ZTF) is a robotic time-domain survey of the northern sky capable of scanning more than 3700 $\deg^2$ $\text{hr}^{-1}$ with its 47 $\deg^2$ field of view, 600 megapixel camera and read out time of just 8 seconds. ZTF can observe to a median $5 \sigma$ r-band depth of 20.6 mag \citep{Graham2019, Bellm2019, Masci2019}. During a full night of observing ZTF may issue as many as 1.2 million alerts ($\sim 1/10$ of what is expected for LSST) to science users through its alert broker system \citep{2019PASP..131a8001P}. Contained within the alerts are the detections of moving objects. In particular, if a previously discovered Solar System object is predicted to be at the location of a difference image point-source, the source can be attributed to the known object \citep{Masci2019}. In its first three months of operations, ZTF submitted $\sim 600,000$ observations of minor planets and has discovered about 320 new objects \citep{Bellm2019}. As only its public survey cadence is designed to emulate the baseline cadence of LSST and produce tracklets, the full ZTF dataset provides a good framework to test THOR on data which otherwise wouldn't be readily useful for minor planet discovery.

\subsection{ZTF Alert Dataset}

We downloaded 15 nights worth of ZTF alerts dating from 2018 September 3 to 2018 September 17. We filtered out static sources and selected all alerts with real bogus value above 0.5 emulating future LSST requirements. This yielded a final count of 255,358 (30.9\%) known object observations that could be matched to objects in the Minor Planet Center's orbit catalog. The remaining 572,188 (69.1\%) alerts are unassociated and therefore unknown. It is likely that amongst the unassociated detections there are undiscovered moving objects in addition to astrophysical transients and false positives (e.g. image subtraction artifacts). 

During the 15 night window of alerts ZTF scanned over 20,000 $\deg^2$ of sky (for comparison the simulated survey had a footprint of 160 $\deg^2$). We divided the sky-plane into 15 by 15 $\deg$ patches of sky and for each patch selected nine average known objects in different bins of $a$ as described in Section \ref{sec:test_orbits}. If no known object exists in a bin of $a$ for a particular patch, no test orbit was used for that bin. A total of 821 test orbits were chosen. 

The Hough transforms and orbit determination for each test orbit were conducted with the same parameters and configuration as in the simulated survey (300 equally binned velocities between -0.1 to 0.1 $\deg \text{ day}^{-1}$ and a maximum allowable clustering radius of 0.005 $\deg$). For clustering, the contamination threshold was set to 20\% and the minimum cluster size was set to five observations. For orbit determination, the $\chi^2_\nu$ and contamination thresholds were set to ($10^5$, 20\%), ($10$, 20\%), and ($10$, 0\%) for IOD, OD, and the iterative combination of differential correction and attribution, respectively. The maximum attribution distance was set to one arcsecond. The minimum number of observations and minimum arc length for an orbit to be accepted was set to five observations and one day, respectively.

Compared to the simulated survey, nearly 100 times as many test orbits were chosen. We grouped the test orbits into chunks of 5 patches (yielding a maximum of 45 test orbits per chunk). Each chunk was submitted to an individual node on the University of Washington's HPC cluster, Hyak. After each chunk finished processing, the recovered orbits were concatenated and a simple algorithm was run to select the orbit with most observations and longest arc for each set of orbits that shared observations. After shared observations were assigned to their best-fitting orbit, a final step of differential correction was executed to update any orbits that had any observations removed.

Like the simulated survey, we consider any known object with at least five detections throughout the 15 nights of alerts to be findable resulting in a total of 21,542 such objects. Assuming a classical MOPS approach we find that only 9,381 (44.8\%) of the known objects would be findable. With the ZMODE algorithm we find 14,291 (68.2\%) would be findable\footnote{The numbers stated here assume {\em ideal} performance by both pipelines -- i.e., 100\% completeness.}. ZMODE is designed to work with ZTF's cadence and attempts to link objects with a minimum of four detections over four night intervals, while the classical LSST MOPS algorithm looks to link three pairs of detections over 15 night-long windows thus at minimum requiring six observations. We consider any object recovered in a pure orbit with at least five observations to be found.

\subsection{THOR Performance}

\begin{figure*}[!htb]
  \centerline{\includegraphics[width=0.60\textwidth,keepaspectratio]{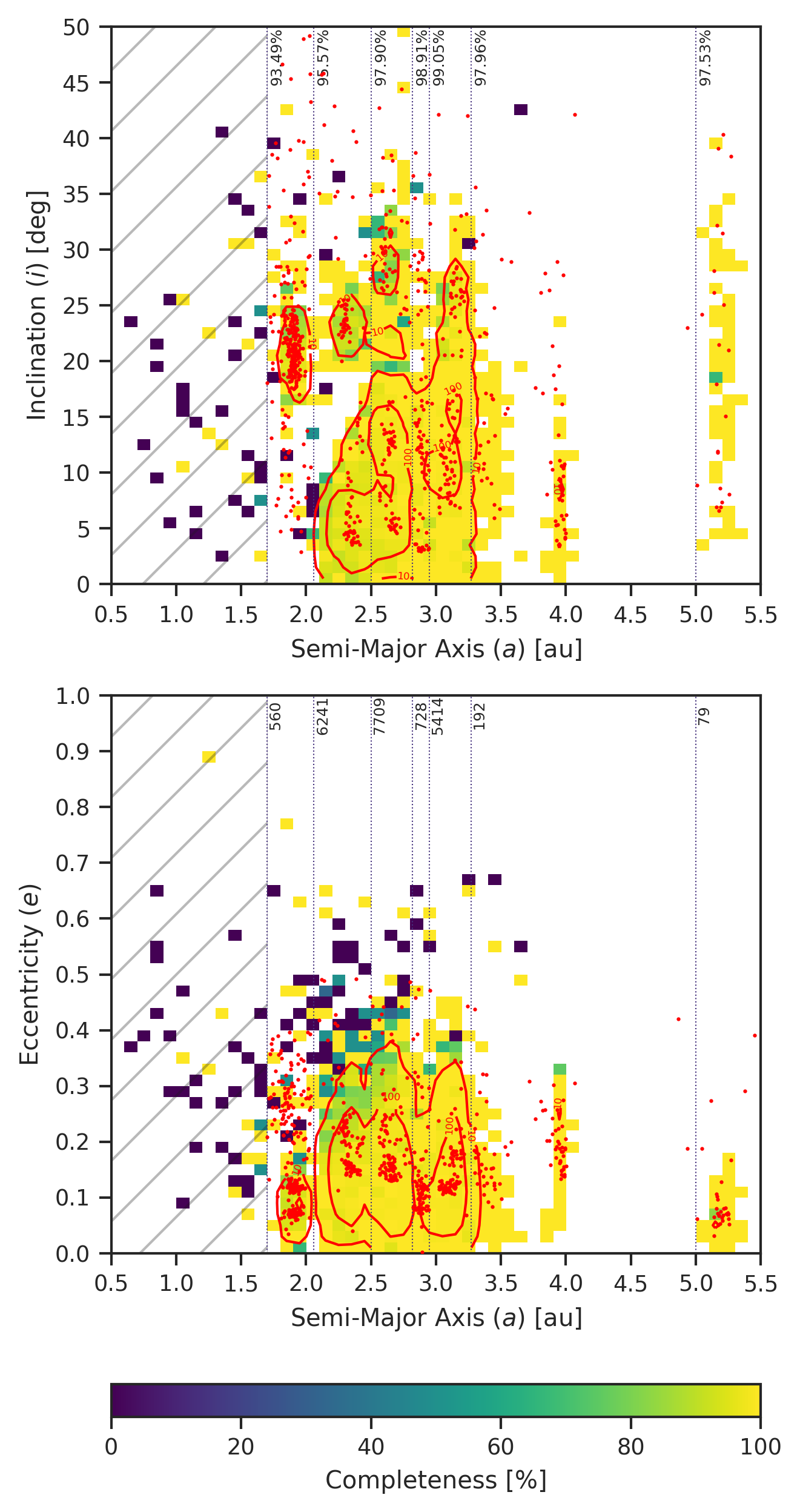}}
  \caption{In the top panel, percentage of objects found in pure or partial clusters is visualized in bins of semi-major axis ($a$) and inclination ($i$). These objects would be linked after an application of an optimal IOD filtering routine followed by orbit determination. Number density contours are drawn as red lines to show the number of objects findable (five or more detections through out the two weeks of ZTF alerts). The hatched area indicates the region of semi-major axis space where no test orbits were chosen. The 821 test orbits used are plotted as red points. The vertical dashed lines indicate the five semi-major axis bin edges, with the overall percentage completeness per bin written at the top. In the bottom panel, percent completeness in bins of semi-major axis ($a$) and eccentricity ($e$) is plotted, contours and test orbits are plotted in the same style as in the top panel. Instead of percent completeness in each of the five bins we now explicitly state the number of objects found. This figure was generated using  \href{https://nbviewer.jupyter.org/github/moeyensj/thor_notebooks/blob/master/paper1/plots_ztf.ipynb}{plots\_ztf.ipynb}.}
  \label{fig:ztf_compeletness}
\end{figure*}

THOR recovered 21,723 candidate orbits consisting of nearly 167,000 unique observations from the described 2-week slice of the ZTF dataset. These include 21,018 pure orbits for 20,940 (97.2\%) of the 21,542 minor planets with at least five detections. The remaining 705 orbits contain either only unassociated/unknown observations or a mix of associated and unnassociated observations. Of the 21,723 recovered orbits, 1,783 orbits contain only singleton observations (orbits made completely without tracklets). In the top panel of Figure \ref{fig:ztf_compeletness} we plot completeness in bins of semi-major axis ($a$) and inclination ($i$). The red number density contours show the number of objects that should be findable while the red points indicate the chosen test orbits. The vertical dashed lines trace the bin edges. The outer Solar System bin stretches from 5.0 to 50 au but plotted in its entirety. The percentages at the top of each bin list the overall completeness in each bin. In the bottom panel, we repeat the same plot configuration but instead of inclination we plot eccentricity ($e$). We replace the percentage completeness values in each bin with the number of found objects to give the percentages greater context. As in the simulated survey, the performance on ZTF alerts beyond 2.5 au suggests most of the findable objects were recovered (98.4\%). The performance in the bins nearer than 2.5 au is better than in the simulated survey. This is likely due to the designed overlap of selecting nine orbits per 15x15 $\deg^2$ patch of sky and setting the cell size to 1000 $\deg^2$ (see Figure \ref{fig:ztf_footprint}). Note that in the bottom panel it is clear that high eccentricity objects were not favored in recovery. Test orbit selection was limited to orbits with eccentricity less than 0.5 and some of these objects may appear non-linearly within the co-rotating frames of the test orbits. Overall completeness for the Main Belt and beyond ($a > 1.7$ au) is 97.4\%, including serendipitiously linking 28.9\% of objects with orbits interior to the Main Belt ($a < 1.7$ au).

Our test orbit selection yielded a total of 821 orbits, however, only 621 of those orbits lead to known objects being found. 220 of the test orbits yielded no known object discoveries, of which 88 had no findable objects in their co-rotating frame of reference. This is due to several reasons: first, some of the selected test orbits might belong to objects which are too dim to be detectable by ZTF. Second, since the orbits are run in series moving outward in semi-major axis space, it is likely that some orbits that ran earlier swept out their population of findable objects resulting in the later test orbits to not have any objects to find especially if the two orbits were similar.

In all, the total number of known objects recovered is 20,940 (97.2\%) of the 21,542 with five or more observations. This represents a factor of $\sim 2$ discovery increase over traditional MOPS, and a factor of $\sim 1.5$ increase over ZMODE.

\subsection{Computational Performance}

We grouped each set of test orbits into chunks of 5 patches. Each group of test orbits was then set to a run on a node on UW's HPC cluster, Hyak. A total of 23 nodes, with 28-cores per node were used. The SLURM jobs were submitted to the backfill (``checkpoint") queue, meaning that jobs are frequently preempted and paused in favor of jobs submitted by the node owners. Regardless of multiple preemptions and job restarts, the full run on two weeks of ZTF took less than 18 hours. Some nodes finished in as little as 10 minutes, the longest running job took just under 16 hours.

\subsection{Completeness and Purity Considerations}

We next characterize the completeness, purity, and any additional filtering required for submission of candidate new discoveries to the Minor Planet Center. In Section \ref{sec:2018_analysis}, we evaluate which objects we would have discovered and reported in 2018, given the state of MPC's orbit catalog at the time. In Section \ref{sec:2021_analysis}, we use a more recent catalog of known orbits from April 2021 to assess the purity of the sample we would have submitted in 2018, and look for still unknown objects.

\subsubsection{2018 Orbit Catalog Analysis}\label{sec:2018_analysis}
The ZTF alerts that correspond to possible observations of known minor planets are labeled as such as part of the ZTF alert pipeline. These labels or associations are the `ground truth' with which we compare THOR's performance. As stated in the previous section, THOR recovered 21,723 candidate orbits including 21,018 pure orbits for 20,940 (97.2\%) of the 21,542 minor planets with at least five detections.

\begin{figure}[hbt!]
  \centerline{\includegraphics[width=0.5\textwidth,keepaspectratio]{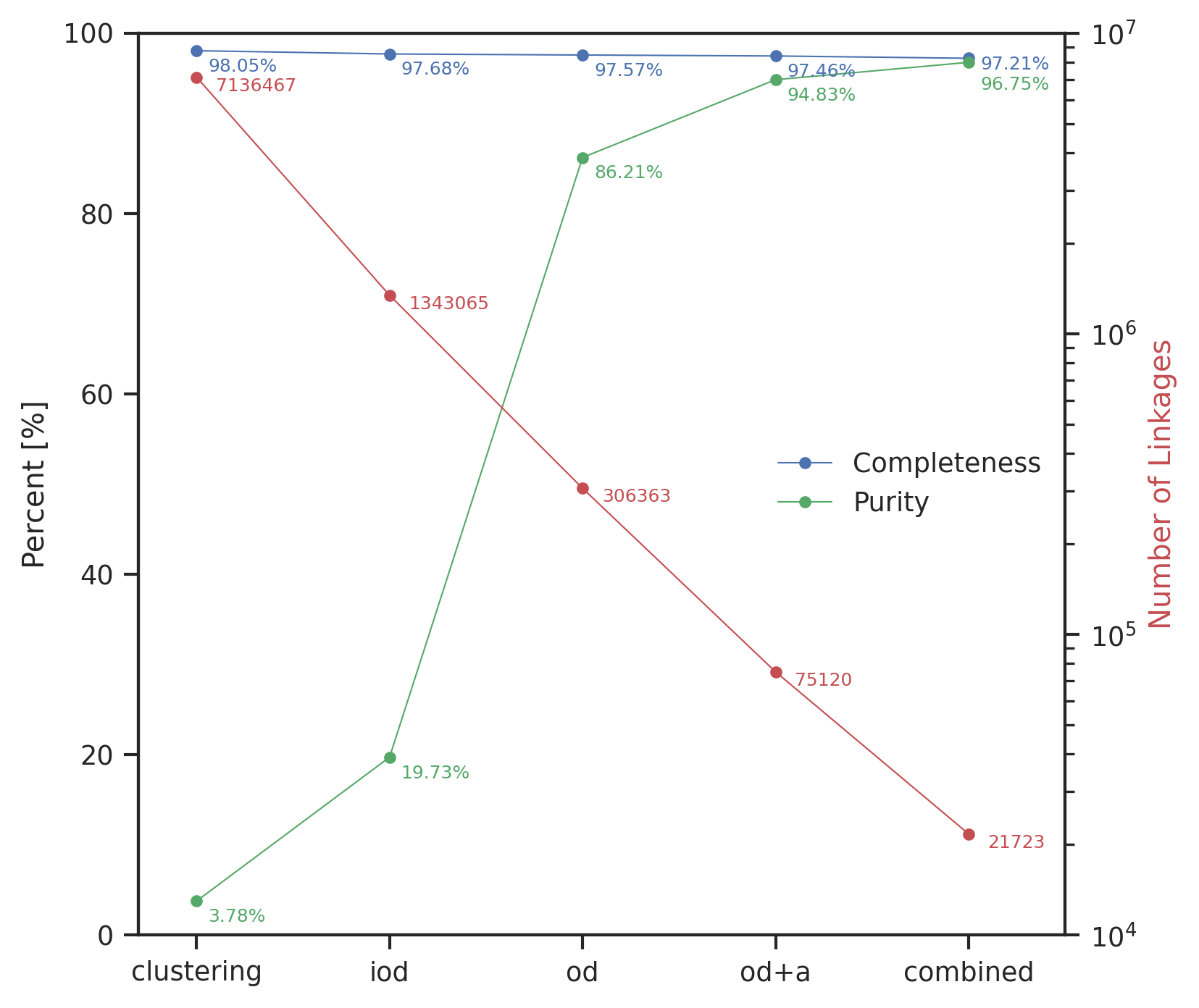}}
  \caption{Completeness and purity as functions of pipeline components are plotted in blue and green, respectively. The number of linkages is plotted in red. The ``combined" step is the simple algorithm used to combine orbits recovered by individual nodes on the HPC cluster. Each subsequent component of THOR enhances linkage purity while completeness remains largely unchanged. The number of linkages quickly approaches the true number of recovered objects. This figure was generated using  \href{https://nbviewer.jupyter.org/github/moeyensj/thor_notebooks/blob/master/paper1/plots_ztf.ipynb}{plots\_ztf.ipynb}.}
  \label{fig:ztf_component_summary_2018}
\end{figure}

In Figure \ref{fig:ztf_component_summary_2018}, we plot completeness and purity as functions of the different pipeline components. The purity\footnote{Note that the purity as computed here is actually a {\em lower limit}. The 705 orbits that have been identified as mixed or unknown will contain objects that were undiscovered in 2018, objects that had known orbits not of sufficient quality such that observations could be attributed to them, as well as truly erroneous linkages.} rises from only 3.8\% at the Hough transform stage to 96.8\% for the final recovered orbits, while -- quite strikingly -- completeness drops by only 0.8\%. This suggests that further completeness is likely to be found by improving our test orbit selection algorithm. In absolute terms, the number of linkages drops from $\sim 7 \times 10^6$ clusters to 21,723 recovered orbits. Among the 21,018 pure orbits are 150 orbits that contain only sub-arcs of the longer observations arcs of 72 unique objects in the dataset. The continued presence of these duplicates suggests further improvements to the orbit merging and extension algorithm are necessary.

We apply two additional filters to produce a final set of high-confidence candidates that could be submitted to the MPC. First, we remove any observations that occurred within 30 minutes of another observation. This helps in removing observations of stationary sources that may not have been filtered out during preprocessing. Doing so reduces the number of unknown orbits from 705 to 526. Second, we remove any orbits that contain the observations of known objects. In practice, if THOR were deployed as a discovery algorithm, it would not run on observations that have already been associated with known objects. This reduces the number of unknown orbits to 488 high quality discovery candidates which, had THOR been running in 2018, would have been submitted to the MPC.

\subsubsection{2021 Orbit Catalog Analysis}\label{sec:2021_analysis}
At the time of writing, two and a half years of minor planet discovery have passed since ZTF made the observations in 2018. We downloaded the Minor Planet Center's (MPC) catalog of known orbits on 2021 April 20, and ran attribution of the MPC's orbits on the two weeks of ZTF observations creating a new set of updated associations. We then analyzed how THOR performed with the benefit of hindsight to establish the lower limits on the purity of the sample that would have been reported to the MPC in 2018.

\begin{figure*}[htb!]
  \centerline{\includegraphics[width=0.95\textwidth,keepaspectratio]{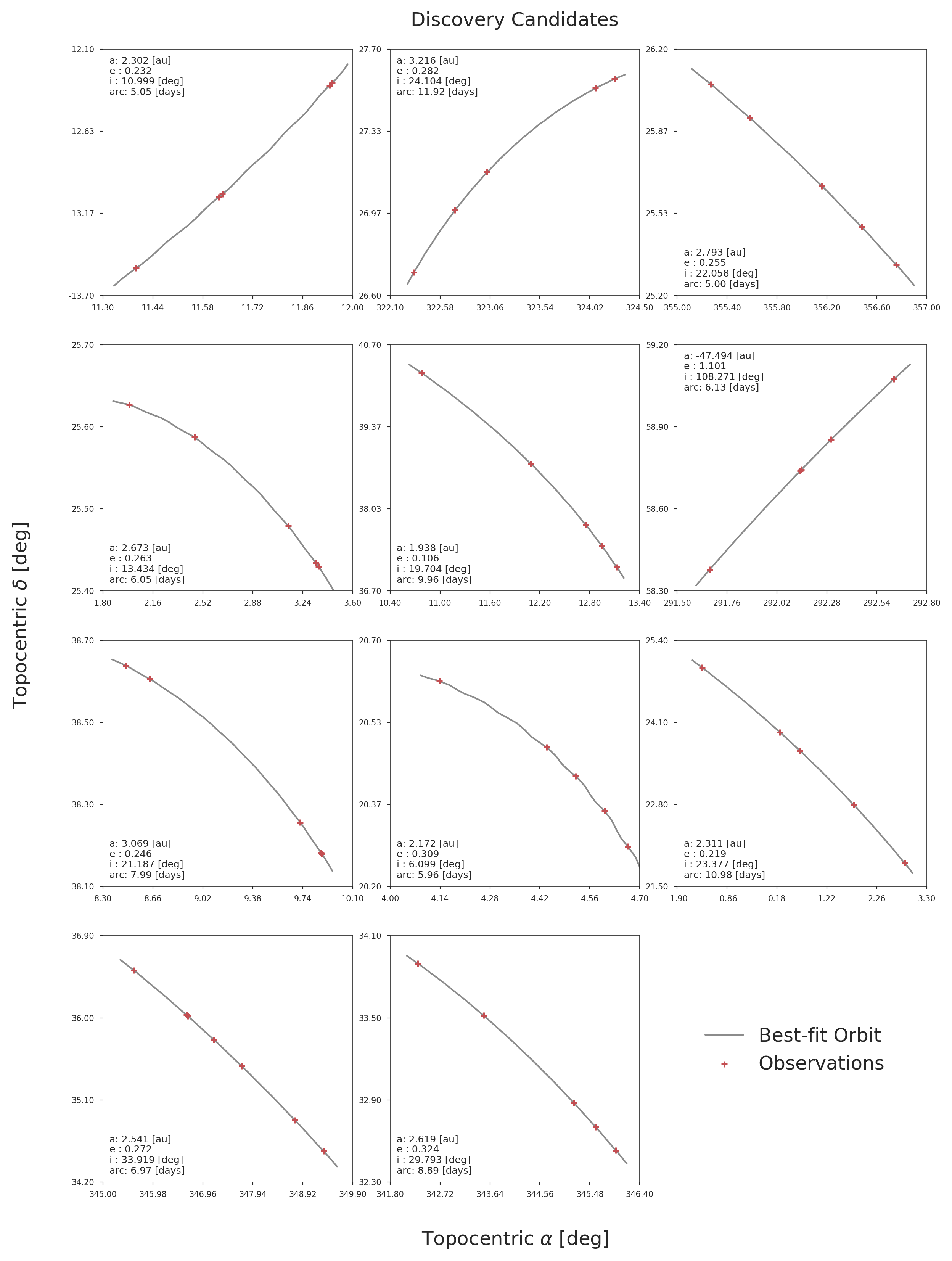}}
  \caption{The observations of the 11 discovery candidates identified in Section \ref{sec:2021_analysis} are plotted in red, with the sky-plane motion of their best-fit orbits plotted as lines. 10 of the objects show MBA-like best fit orbit solutions. The remaining object has a hyperbolic orbit solution and corresponds to recovery observations of the hyperbolic comet C/2018 U1. The ``wiggles'' apparent in some of the best-fit orbit lines are due to the motion of the observer (topocentric motion). This figure was generated using  \href{https://nbviewer.jupyter.org/github/moeyensj/thor_notebooks/blob/master/paper1/plots_ztf.ipynb}{plots\_ztf.ipynb}.}
  \label{fig:discovery_candidates}
\end{figure*}

Of the 488 orbits that were identified as high quality, submission-ready, discoveries using the 2018 associations, we find that 477 correspond to observations of 476 unique minor planets that have been discovered or had their orbits improved since 2018 (one object was identified twice, as two shorter sub-arcs). The remaining 11 orbits in the high quality sample are as yet undiscovered objects or false linkages; based on visual inspection and comparisons of their magnitudes they do appear to be real objects. The observations and predicted motion of the 11 discovery candidates are plotted in Figure \ref{fig:discovery_candidates}.

Using the IMCCE SkyBot service \citep{SkyBot}, we conducted a cone-search for all observations belonging to the 11 orbits. The observations of 10 candidates could not be associated to any known objects within 10 arseconds. Only the candidate with $e > 1$ orbit had all observations within 5~arcsec of a known object -- the comet C/2018 U1. C/2018 U1 is classified as a hyperbolic comet and was discovered on 2018 October 27. We compared the orbit fit of C/2018 U1 to the orbit as reported in JPL's SBDB\footnote{\url{https://ssd.jpl.nasa.gov/sbdb.cgi}}. The inclinations agree to within 0.3 deg, the perihelion distances agree to within 0.2 au, while our fit overestimates eccentricity by 0.1. Given the short arc (6.13 days) and small number of observations (5 observations) with which the orbit was fit, we conclude our solution agrees remarkably well with the one present in the SBDB. The 2-week slice of ZTF observations were made 6-8 weeks prior to the discovery date. Had THOR been running at the time when data was taken, this comet would have been discovered by ZTF.

We, therefore, conclude that the purity of submissions -- an important consideration for the Minor Planet Center -- is at least 97.7\%. If all 11 discovery candidates are confirmed as real, then the purity of the submitted sample could be as high as 100\%. This demonstrates that, at a practical level, THOR is ready to be used for discovery and reporting of new discoveries with a high degree of confidence.

\section{Summary and Future Work}\label{sec:discussion}

We present an observer- and cadence-independent asteroid linking algorithm, which we name ``Tracket-less Heliocentric Orbit Recovery" (THOR), and show its performance on both simulated and real detections. We have shown that THOR can link Main Belt asteroids and more distant populations at high completeness and at moderate computational cost. On two weeks of simulated data, THOR recovered 91.3\% of the 18,332 that were ideally findable, whereas a tracklet-based linking algorithm would have recovered none. On two weeks of ZTF data, THOR linked 97.2\% of the 21,542 objects with at least five detections (a factor of $\sim2$ recovery increase over MOPS and a factor of $\sim1.5$ increase over ZMODE). THOR recovered orbits for 97.4\% of objects beyond 1.7 au, with 98.4\% of objects recovered beyond 2.5 au. Furthermore, by comparing the 2018 sample to the catalog of orbits as presently known (April 2021), we show that the lower limit on purity of THOR submissions to the MPC would be 97.7\% and -- assuming all candidates shown in Figure~\ref{fig:discovery_candidates} are confirmed as real -- possibly as high as 100\%. This, in combination with its capability to discover objects regardless of cadence or observer, renders it immediately useful for Main Belt and KBO searches on survey data and archival datasets. 

While THOR can be applied to running surveys, application to archival datasets -- such as ZTF, CSS, or PanSTARRS -- is interesting as well. The most straightforward way to do this would be to run a sliding $\sim$two week window over the dataset, from beginning to the end, running THOR at each window instance. For each window run, orbits could be computed for the discovered objects and projected to the past (and future) for discovery of additional observations (and improvement of orbital solutions). This is the approach we ourselves plan to take with the ZTF archival data. Assuming the 11 objects discovered here are representative of remaining undiscovered objects in ZTF, this search would likely yield on order of $1,000$ asteroids. The discovery potential with deeper archival datasets (e.g., DES data or the DECam archive) is likely to be significantly larger.

\subsection{Future Directions}

There are several avenues for improvement which we discuss here below.
\\

{\bf Test orbit selection:} As discussed previously, THOR can discover objects at very high completeness in the Main Belt and beyond using the current test orbit selection criteria. Those selection criteria are, however, not ideal for the inner Main Belt, NEOs, objects interior to Earth's orbit or objects on generally unusual orbits (e.g. ISOs). For the linking algorithm to be universally viable, these populations need to be reliably discoverable. The focus of this paper has been to demonstrate that the overall approach works even with a rather simplistic orbit selection algorithm; developing an optimal test orbit allocation algorithm is the focus of follow-up work and is currently in progress.
\\

{\bf Hyperparameter optimization:} Throughout the description of the algorithm's performance on both the simulated survey and on two weeks of ZTF alerts we stated the values of the hyperparameters used. These values were decided on after numerous trial runs, early testing, and optimizations. We plan to quantify in much greater detail how THOR performs as these parameters change, and how these parameter might change with different combinations of datasets and surveys. This includes automated data-driven selection of the clustering radius, the velocity grid and its granularity, as well as decisions on the width of the time window over which to attempt linking.
\\

{\bf Orbit determination:} With our goal of extending THOR to the NEO population will likely come the need to make improvements to the orbit determination component of THOR. Improvements such as iteration at the IOD stage as opposed to delegating the work directly to the differential correction could be useful here. We have also started researching the possibility of bootstrapping orbit determination directly from the co-rotating reference frame of the test orbit. The test orbit effectively serves the initial orbit guess for the orbits of objects contained in the tangent plane centered on the motion of the test orbit. The relative velocities (determined by the Hough transform), and the location in the tangent plane, should adequately constrain four of six orbital parameters. The remaining two can be fit for at lower computation cost than performing initial orbit determination and differential correction on sky-plane coordinates.
\\

{\bf Scalability and ``discovery-as-a-service":} We plan to deploy THOR as a scalable cloud-based service to which observations can be submitted for automated identification of Solar System objects with candidate discoveries returned to the user and/or submitted to the Minor Planet Center. This would effectively provide the Solar System community with a \textit{``discovery-as-a-service''} platform. The advantage of such a cloud-based platform is that it can elastically scale to meet THOR's computational needs as datasets are submitted. Additionally, a centralized discovery service would enable cross-survey and cross-dataset discovery.
\\

\subsection{Software and Transparency}

The THOR code is maintained on GitHub. All the analysis presented in this document, the figures, and the datasets are available online\footnote{\url{https://github.com/moeyensj/thor}}. Of particular note is a series of Jupyter notebooks that reproduce all the numbers, results, and plots in this paper\footnote{\url{https://github.com/moeyensj/thor_notebooks}}. Included with the GitHub repository are instructions on how to install all the software dependencies and code in a containerized environment using Docker. We encourage the reader to visit the GitHub repository for more details.

\acknowledgements

J. Moeyens wishes to acknowledge the support of the Asteroid Institute and B612 Foundation. This research was made possible by major gifts to support the ADAM project provided by the W.K. Bowes Jr. Foundation, Steve Jurvetson, The P. Rawls Family Fund, Tito's Handmade Vodka, Yishan Wong and Kimberly Algeri-Wong, and two anonymous donors, along with Founding Circle and Asteroid Circle members: B. Anders, R. Armstrong, G. Baehr, The Barringer Crater Company, B. Burton, D. Carlson, S. Cerf, V. Cerf, C. Chapman, Y. Chapman, J. Chervenak, D. Corrigan, E. Corrigan, A. Denton, E. Dyson, A. Eustace, A. Fritz, L. Fritz, S. Galitsky, E. Gillum, L. Girand, Glaser Progress Foundation, D. Glasgow, J. Grimm, S. Grimm, G. Gruener, V. K. Hsu \& Sons Foundation Ltd., J. Huang, J. D. Jameson, J. Jameson, M. Jonsson Family Foundation, D. Kaiser, S. Krausz, V. Lašas, J. Leszczenski, D. Liddle, S. Mak, G.McAdoo, S. McGregor, J. Mercer, M. Mullenweg, D. Murphy, P. Norvig, S. Pishevar, R. Quindlen, N. Ramsey, R. Rothrock, E. Sahakian, R. Schweickart, A. Slater, T. Trueman, F. B. Vaughn, R. C. Vaughn, M. Welty, S. Welty, B. Wheeler, M. Wyndowe, and seven anonymous donors in addition to donors from over 46 countries around the world. J. Moeyens further wishes to thank the LSSTC Data Science Fellowship Program; his time as a Fellow has benefited this work.

M. Juri\'{c} and J. Moeyens wish to acknowledge the support of the Washington Research Foundation Data Science Term Chair fund, and the University of Washington Provost’s Initiative in Data-Intensive Discovery. 

All authors acknowledge support from the University of Washington College of Arts and Sciences, Department of Astronomy, and the DIRAC Institute. The DIRAC Institute is supported through generous gifts from the Charles and Lisa Simonyi Fund for Arts and Sciences, and the Washington Research Foundation.

Based on observations obtained with the Samuel Oschin Telescope 48-inch and the 60-inch Telescope at the Palomar Observatory as part of the Zwicky Transient Facility project. ZTF is supported by the National Science Foundation under Grant No. AST-1440341 and a collaboration including Caltech, IPAC, the Weizmann Institute for Science, the Oskar Klein Center at Stockholm University, the University of Maryland, the University of Washington, Deutsches Elektronen-Synchrotron and Humboldt University, Los Alamos National Laboratories, the TANGO Consortium of Taiwan, the University of Wisconsin at Milwaukee, and Lawrence Berkeley National Laboratories. Operations are conducted by COO, IPAC, and UW.

This work was supported by the Washington Research Foundation and by a Data Science Environments project award from the Gordon and Betty Moore Foundation (Award \#2013-10-29) and the Alfred P. Sloan Foundation (Award \#3835) to the University of Washington eScience Institute.

This research has made use of data and/or services provided by the International Astronomical Union's Minor Planet Center.

This work was facilitated through the use of advanced computational, storage, and networking infrastructure provided by the Hyak supercomputer system and funded by the STF at the University of Washington.

The authors would like to thank Victoria Meadows, Matthew McQuinn, and Scott Anderson at the University of Washington for their many helpful comments on this work. 

\software{oorb \citep{oorb}, spiceypy \citep{Annex2020}, numpy \citep{numpy}, pandas \citep{pandas}, scikit-learn \citep{scikit-learn}, astropy \citep{astropy-1, astropy-2}, matplotlib \citep{matplotlib}, seaborn \citep{seaborn}, plotly \citep{plotly}, bokeh \citep{bokeh}}

The authors are grateful to the anonymous referee for a thorough review of multiple revisions of this manuscript, and for encouraging us to complete and add the orbit determination elements which were originally indented to be a part of a follow-up paper. This has enabled us to deliver a fully-functional software package that can be immediately applied for Solar System object searches.

\appendix
\section{Transformations, Rotations and Projections}\label{sec:appendix}

Here we summarize the key equations and relations used to carry out the heliocentric transformation and projection into the co-rotating frame of the test orbit.

\subsection{Equatorial and Ecliptic Coordinates}\label{sec:coordinates}

For convenience, we reproduce a subset of the equations presented in \cite{Bernstein2000} that outline the coordinate transformations between the equatorial and ecliptic systems. A set of observations in equatorial coordinates $(\alpha, \delta)$ can be transformed into Cartesian equatorial coordinates as follows

\begin{equation}\label{eq1}
    \begin{bmatrix}
        x \\ 
        y \\ 
        z
    \end{bmatrix}_{eq} = \begin{bmatrix}
        \cos \delta \cos \alpha \\ 
        \cos \delta \sin \alpha \\ 
        \sin \delta 
    \end{bmatrix}.
\end{equation}

Similarly, a set of ecliptic coordinates $(\lambda, \beta)$ can be transformed into Cartesian ecliptic coordinates

\begin{equation}\label{eq2}
    \begin{bmatrix}
        x \\ 
        y \\ 
        z
    \end{bmatrix}_{ec} = \begin{bmatrix}
        \cos \beta \cos \lambda \\ 
        \cos \beta \sin \lambda \\ 
        \sin \beta 
    \end{bmatrix}.
\end{equation}

Cartesian equatorial coordinates can be transformed to ecliptic Cartesian coordinates. Here $\epsilon$ is the obliquity of the ecliptic ($\epsilon = 84381.448 ''$)

\begin{equation}\label{eq3}
    \begin{bmatrix}
        x\\ 
        y\\ 
        z
    \end{bmatrix}_{ec} = \begin{bmatrix}
        1 & 0 &  0\\ 
        0 &  \cos \epsilon  & \sin \epsilon \\ 
        0 &  -\sin \epsilon & \cos \epsilon
    \end{bmatrix} \begin{bmatrix}
        x\\ 
        y\\ 
        z
    \end{bmatrix}_{eq}.
\end{equation}

Likewise, ecliptic Cartesian coordinates can be converted to equatorial Cartesian coordinates

\begin{equation}\label{eq4}
    \begin{bmatrix}
        x \\ 
        y \\ 
        z
    \end{bmatrix}_{eq} = \begin{bmatrix}
        1 & 0 &  0\\ 
        0 &  \cos \epsilon & -\sin \epsilon \\ 
        0 &  \sin \epsilon & \cos \epsilon
    \end{bmatrix} \begin{bmatrix}
        x \\ 
        y \\ 
        z
    \end{bmatrix}_{ec}.
\end{equation}

Using equations \ref{eq1} - \ref{eq4}, coordinates in either the angular or Cartesian form for both the ecliptic and equatorial systems can be trivially transformed between the respective systems and the different forms. To proceed with the heliocentric transformation which is described in the next section, THOR transforms all equatorial coordinates to ecliptic coordinates. 

\subsection{Heliocentric Transformation}\label{sec:heliocentric_tranform}

Before proceeding, it is useful to define certain quantities in relation to the assumed observing geometry. The location of the observer in the heliocentric frame we denote as $\vec{r}_o$ and is assumed to be known. The distance from the Sun to the observer is then simply the magnitude of said vector ($\left | \vec{r}_o \right | = r_o$). The unit vector $n_{ao}$ lies in the direction of an observed asteroid from the position of the observer. Unless the asteroid's orbit is well-constrained and known beforehand we do not know the distance from the observer to the asteroid. We express the observer to asteroid position vector as $\vec{r}_{ao}$ which has magnitude $\Delta$ and is equivalent to $n_{ao} \cdot \Delta$. Lastly, we denote the unknown heliocentric position vector of the asteroid as $\vec{r}_a$ which has magnitude $r$. 

A fiducial test orbit has heliocentric position vector $\vec{r}$ and velocity $\vec{v}$ given at some epoch, $t$. The location in space and time of an object on the test orbit can be precisely predicted. Assuming there are other minor planets on orbits similar to the test orbit we can propagate the test orbit to all times where it could have been detected in a survey and gather actual detections near its predicted location (Sections \ref{sec:testorbit} - \ref{sec:propagate}). Nearby detections likely belong to moving objects on similar orbits. If we assume those lie at the same heliocentric distance from the Sun, we have enough information to fully constrain the observing geometry for each individual detection allowing the unknown quantities ($\Delta$ and $\vec{r}_a$) to be calculated.

From our assumed observing geometry, we know the direction $\hat{n}_{ao}$ for every gathered detection. Assuming each detection lies at the distance $r$ of the test orbit, we can then calculate $\Delta$:

\begin{equation}\label{eq5}
    \Delta = -\vec{n}_{ao} \cdot \vec{r}_o + \sqrt{(\vec{n}_{ao} \cdot \vec{r}_o ) + r^2 - r_o^2}.
\end{equation}

Once we know $\Delta$ we can solve for $\vec{r}_{a}$, where 

\begin{equation}\label{eq6}
    \vec{r}_{a} = \vec{r}_{ao} + \vec{r}_o.
\end{equation}

At this point, we can attribute a heliocentric position vector to every gathered detection at its specific epoch, again under the assumption that the detection belongs to an orbit similar to the test orbit.

\subsection{Transformation into Frame of the Test Orbit}\label{sec:orbit_transform}

Once the heliocentric coordinates of each observation have been calculated, the observations of other minor planets can be transformed into the frame of the test orbit. This is accomplished using a double rotation. First, the detections are rotated so that the test orbit lies in the x-y plane. Second, the detections are rotated so the test orbit lies along the x axis. 

To perform the first rotation which we denote as $\vec{R}_1$, we transform the detections so that $\vec{r}$ lies in the x-y plane. This is equivalent to rotating the vector normal to the plane of the orbit ($\hat{n}$) towards the z-axis as defined in heliocentric space ($\hat{z}$). We now follow the method outlined in \citep{rotationTwoUnitVectors}. The rotation axis, $\hat{v}$, can be found by taking the cross product of $\hat{n}$ and $\hat{z}$,
\begin{equation}
    \hat{v} = \hat{n} \times \hat{z}.
\end{equation}
The rotation matrix, $\vec{R}_1$, is then
\begin{equation}
    \vec{R}_1 = I + [\hat{v}]_\times + [\hat{v}]_\times^2 \frac{1}{1 + c},
\end{equation}
where $c = \hat{n} \cdot \hat{z}$ and $[\hat{v}]_\times$ is the skew-symmetric cross product matrix of $\hat{v}$,
\begin{equation}
    [\hat{v}]_\times \equiv \begin{bmatrix}
    0 & -\hat{v}_3 & \hat{v}_2 \\ 
    \hat{v}_3 & 0 & -\hat{v}_1 \\ 
    -\hat{v}_2 & \hat{v}_1 & 0
    \end{bmatrix}.
\end{equation}
Finally, we need to rotate the detections so that $\vec{r}$ lies along the x-axis. We denote this rotation as $\vec{R}_2$. Let $\hat{r}'$ be a unit vector in the direction of a detection or test orbit rotated into the x-y plane by $\vec{R}_1$. The angle of rotation towards the x-axis, $\alpha$, is equivalent to $-\hat{x} \cdot \hat{r}'$, 
\begin{equation}
    \vec{R}_2 = \begin{bmatrix}
        \cos \alpha & \sin \alpha & 0 \\
        -\sin \alpha & \cos \alpha & 0 \\
        0 & 0 & 1
    \end{bmatrix}.
\end{equation}
The complete rotation matrix is $\vec{M} = \vec{R}_2 \cdot \vec{R}_1$. 

\subsection{Gnomonic Projection}\label{sec:gnomonic}

After the observations have been rotated, they can be projected into a gnomonic tangent plane \citep{gnomonicProjection}. The gnomonic tangent plane preserves great circle distances and so offers a natural projection in which to discover minor planets relative to a test orbit. The gnomomic projection is also used by both \cite{Holman2018} and \cite{Bernstein2000}. The gnomonic projection can be accomplished using either spherical or Cartesian coordinates. For completeness, we list both sets of equations. In our implementation we directly project the transformed and rotated Cartesian coordinates onto the gnomonic tangent plane.

To project a set of spherical coordinates $(\lambda, \phi)$ onto a gnomonic plane ($\theta_X, \theta_Y$) tangent about point $(\lambda_0, \phi_0)$ (in our case, the location of the test orbit at each epoch in which detections were found and gathered) 
\begin{equation}
    \theta_X = \frac{\cos\phi \sin(\lambda - \lambda_0)}{\sin \phi_0 \sin \phi + \cos \phi_0 \cos \phi \cos (\lambda - \lambda_0)},
\end{equation}

\begin{equation}
    \theta_Y = \frac{\cos\phi_0 \sin \phi - \sin \phi_0 \cos \phi \cos(\lambda - \lambda_0)}{\sin \phi_0 \sin \phi + \cos \phi_0 \cos \phi \cos (\lambda - \lambda_0)}.
\end{equation}
The same projection can be achieved using the rotated Cartesian coordinates described in the previous section. In this case, by convention, the test orbit lies at $(\lambda_0, \phi_0) = (0, 0)$ making the projection simply

\begin{equation}
    \theta_X = \frac{x''}{y''},
\end{equation}

\begin{equation}
    \theta_Y = \frac{z''}{x''}.
\end{equation}
The double prime ($''$) is used to denote the transformed Cartesian coordinates after they have been rotated as described in Section \ref{sec:orbit_transform}. After completing the procedure outlined in Sections \ref{sec:coordinates} - \ref{sec:gnomonic}, we now have a set of detections and their locations in the co-rotating reference frame of a test orbit. From this reference frame, other orbits can be recovered as described in Section \ref{sec:hough}.

\bibliography{thor}{}
\end{document}